\documentclass[aps,prd,showpacs,onecolumn,notitlepage,superscriptaddress,amsmath,amsfont,amssymb,graphicx,10pt]{revtex4-1}
\usepackage{color,graphicx,epsfig}
\usepackage{amsmath} 
\usepackage{amssymb} 
\usepackage{bm}
\usepackage{color}
\usepackage{braket}

\usepackage[english]{babel}

\newcommand {\E}[1]{\times 10^{#1}} 
\newcommand {\e}[1]{\mathrm{~#1}} 
\newcommand{\mc}[1]{\mathcal{#1}} 

\newcommand{\mrm}[1]{\mathrm{#1}}
\newcommand{\ddbar}[0]{\mbox{$D^0$--$\bar D^0$} }

\definecolor{Red}{rgb}{1.,0.,0.}

\begin{document}

\title{Light Colored Scalar as Messenger of Up-Quark Flavor
  Dynamics\\ in Grand Unified Theories}

\author{Ilja Dor\v sner} \email[Electronic address:]{ilja.dorsner@ijs.si}
\affiliation{Department of Physics, University of Sarajevo, Zmaja od Bosne 33-35, 71000
  Sarajevo, Bosnia and Herzegovina}

\author{Svjetlana Fajfer} 
\email[Electronic address:]{svjetlana.fajfer@ijs.si} 
\affiliation{Department of Physics,
  University of Ljubljana, Jadranska 19, 1000 Ljubljana, Slovenia}
\affiliation{J. Stefan Institute, Jamova 39, P. O. Box 3000, 1001 Ljubljana, Slovenia}

\author{Jernej F. Kamenik} 
\email[Electronic address:]{jernej.kamenik@ijs.si} 
\affiliation{J. Stefan Institute, Jamova 39, P. O. Box 3000, 1001 Ljubljana, Slovenia}

\author{Nejc Ko\v snik} 
\email[Electronic address:]{nejc.kosnik@ijs.si} 
\affiliation{J. Stefan Institute, Jamova 39, P. O. Box 3000, 1001 Ljubljana, Slovenia}

\date{\today}

\begin{abstract}
  The measured forward-backward asymmetry in the $t \bar t$ production
  at the Tevatron might be explained by the additional exchange of a
  colored weak singlet scalar. Such state appears in some of the grand
  unified theories and its interactions with the up-quarks are purely
  antisymmetric in flavor space. We systematically investigate the
  resulting impact on charm and top quark physics. The constraints on
  the relevant Yukawa couplings come from the experimentally measured
  observables related to $D^0$--$\bar D^0$ oscillations, as well as
  di-jet and single top production measurements at the Tevatron. After
  fully constraining the relevant Yukawa couplings, we predict
  possible signatures of this model in rare top quark decays. In a
  class of grand unified models we demonstrate how the obtained
  information enables to constrain the Yukawa couplings of the
  up-quarks at very high energy scale.
\end{abstract}

\pacs{14.65.Ha,12.10.-g,14.40.Lb}

\maketitle

\section{Introduction}
During the last decade, rare processes of down-type quarks have been
proven instrumental in the study of physics beyond the Standard
Model~(SM). On the other hand, the discovery of novel effects in the
up-quark sector is still often dismissed as very unlikely. In the SM
in particular, rare charm processes are mostly dominated by long
distance dynamics, the short distance contributions being subject to
severe Glashow-Iliopoulos-Maiani cancellations. This is the case in
\ddbar oscillations as well as in the $c \to u \gamma$ and $c \to u
l^+ l^-$ decays.  Nonetheless, the experimental situation in the charm
sector is already very restrictive and present measurements constrain
many models of new
physics~\cite{Golowich:2009ii,Falk:2004wg,Gedalia:2009kh,Bigi:2009df,Grossman:2006jg,Bigi:2009aw,Petrov:2010gy}.

The heaviest up-type quark, i.e., the top quark, has been carefully
investigated since its discovery in 1995. Although enormous progress
has been achieved, the prospects for observing flavor changing neutral
current mediated top quark decays at the Large Hadron Collider~(LHC)
or Tevatron are rather weak. The LHC experiments might be sensitive to
the branching fractions of the order of $10^{-5}$ or less. However, the
majority of well studied models of new physics~(NP) predict much weaker
signals. It thus seems that the search for NP in the top
sector should be directed towards other
processes~\cite{Bernreuther:2008ju,AguilarSaavedra:2010rx}.

The recent CDF measurement of a large positive forward-backward
asymmetry~(FBA) in the $t \bar t$ production disagrees with the SM
prediction and thus indicates a possible presence of NP.  Among the
many proposed NP scenarios aiming to explain this discrepancy, we have
recently suggested~\cite{Dorsner:2009mq} a theoretically
well-motivated $SU(5)$ grand unified theory~(GUT)~\cite{Georgi:1974sy}
model~\cite{Perez:2007rm}. The model has an appealing feature of
correlating the presence of light colored scalars stemming from a
$45$-dimensional Higgs representation with bounds on the proton
lifetime. Namely, the aforementioned representation contains among
other states two colored scalars---$\Delta_6=(\bar 3,1,4/3)$ and
$\Delta_1=(8,2,1/2)$---that have masses below or of the order of
$1$\,TeV when partial proton decay lifetimes are predicted to be at,
or slightly above, the current experimental bounds.  In this regime
$\Delta_6$ can help to reconcile the SM theoretical prediction of the
forward-backward asymmetry in $t\bar t$
production~\cite{Arnold:2009ay,Shu:2009xf,Dorsner:2009mq}, which is
more than $2\sigma$ below the measured value, while preserving the
agreement in the total $t \bar t$ production cross section. On the
other hand, the contributions from $\Delta_1$ are required to be
suppressed~\cite{Dorsner:2009mq}.

Relatively light colored scalars may also also arise in other
extensions of the SM. Generically, lower bounds of their masses can be
inferred from their strong interaction mediated production at hadron
colliders. In addition however, their couplings to SM matter fields
will induce contributions to flavor observables.  We investigate
constraints and predictions of flavor observables in the up-quark
sector in the presence of a color triplet weak singlet scalar with
hypercharge $4/3$. Furthermore, we address the resulting constraints
on the Yukawa sector of a particular class of GUT models that employ
the $45$-dimensional $SU(5)$ Higgs representation containing such
state~\cite{Georgi:1979df}. As has been noticed recently, $\Delta_6$
exchange does not contribute to $d=6$ proton decay
operators~\cite{Dorsner:2009cu} and can thus be light.

In Section~\ref{framework}, we describe our framework. The \ddbar
constraints are studied in Section~\ref{DDbar}. Constraints
coming from the Tevatron single top and di-jet production cross-section
measurements are presented in Section~\ref{g12}. Resulting predictions
for rare top decays are given in Section~\ref{topdecays}, while
implications of phenomenologically deduced constraints for the entire
up-quark Yukawa sector in GUT models are addressed in
Section~\ref{upyukawa}. Discussion and summary are presented in
Section~\ref{summary}.

\section{Framework}
\label{framework}
The color triplet scalar we study appears in theoretically
well-motivated class of $SU(5)$ models. Namely, it is a part of a
$45$-dimensional Higgs representation, which has been frequently used
in GUT model building to accommodate known fermion masses, improve
unification of gauge couplings, and address proton decay
constraints~\cite{Georgi:1979df,Babu:1984vx,Giveon:1991zm,Dorsner:2006dj,Dorsner:2007fy,Perez:2007rm}. Recall
that matter of the SM is assigned to a $10$- and $5$-dimensional
$SU(5)$ representations, i.e. $\bm{10}_i = (1,1,1)\oplus (\bar
3,1,-2/3) \oplus (3,2,1/6)$ and $\bar{\bm 5}_i = (1,2,-1/2)\oplus(\bar
3,1,1/3)$ where $i=1,2,3$ denotes generation index. It couples to the
$45$-dimensional Higgs representation $\bm{45}$ through the following
operators
\begin{equation}
  V_{45}^\mrm{matter} = (Y_1)^{ij} (\bm{10}^{\alpha\beta})_i (\bar{\bm{5}}_\delta)_j \bm{45}^{*\delta}_{\alpha\beta}\label{M_D&M_E}+(Y_2)^{ij} \epsilon_{\alpha\beta\gamma\delta\epsilon} (\bm{10}^{\alpha\beta})_i (\bm{10}^{\zeta\gamma})_j \bm{45}_\zeta^{\delta\epsilon}.
\end{equation}
Lepton and baryon number violating Yukawa couplings of $\Delta_6$ in
the interaction basis are
\begin{align}
  \mc{L}_{\Delta_6} = \sqrt{2}& [(Y_2)_{ij}-(Y_2)_{ji}]\epsilon_{abc} \bar{u}_{ia} P_L u^c_{jb} \Delta_6^c  + (Y_1)^{ij} \bar{e}_{i} P_L d^c_{ja} \Delta_6^{a*}+ \mrm{H.c.},
\end{align}
where $a$, $b$, $c$ are color indices, while $i$, $j$ denote flavors. Note that the antisymmetric nature of the $\Delta_6$ coupling to the up-quark sector in flavor space is dictated by group theory.
Yukawa couplings of $\Delta_6$ to diquarks in the physical basis,
i.e., the up-quark mass eigenstate basis, acquire a unitary rotation ($U_R$),
which however does not spoil the antisymmetry in flavor
indices. Hence, we define
\begin{equation}
\label{g6}
g_6^{ij} \equiv 2\sqrt{2} \left[U_R^\dagger (Y_2 - Y_2^T) U_R^*\right]^{ij}, \qquad g_6^{ij} = -g_6^{ji},
\end{equation}
altogether with three independent parameters: $g_6^{12}$,
$g_6^{23}$, and $g_6^{13}$. It is the antisymmetry of $g_6$ that is
responsible for the absence of the $d=6$ operators due to $\Delta_6$
exchange that would otherwise contribute to proton
decay~\cite{Dorsner:2009cu}. In what follows we focus our attention on
phenomenological constraints of these entries. 

Note that a tree-level exchange of $\Delta_6$ contributes to $t\bar t$
production cross-section in the $u$-channel. It was emphasized
in~\cite{Dorsner:2009mq} that $\Delta_6$ at and below $1$\,TeV can
enhance the SM prediction of the forward-backward asymmetry
$A_{FB}^{t\bar t}$ while not altering the production cross-section
$\sigma_{t\bar t}$. This has been achieved by finding the parameter
space of $m_{\Delta_6}$ and coupling $g_6^{13}$, contributing to
partonic subprocess $u\bar u \to t\bar t$, in $p\bar p$ collisions,
which together with SM reproduce the measured values of $A_{FB}^{t\bar
  t}$ and $\sigma_{t\bar t}$. The region where experimental
constraints can be satisfied within $1\sigma$ roughly corresponds to a
region, where mass of $\Delta_6$ and the coupling $g_6^{13}$ are
correlated as
\begin{equation}
  \label{eq:g13fit}
  |g_6^{13}| = 0.9(2) + 2.5(4) \frac{m_{\Delta_6}}{1\e{TeV}}.
\end{equation}
We investigate the bounds on the remaining two couplings ($g_6^{12}$,
$g_6^{23}$) in the following sections.


\section{\ddbar mixing constraints}
\label{DDbar}
Mixing of neutral charm mesons is sensitive to interactions which
change charm flavor by 2 units. Scalar $\Delta_6$, if light enough,
may strongly contribute to effective $|\Delta C|=2$ interactions and
expose its Yukawa couplings to the up-type quarks in \ddbar mixing.
In this section we study the contribution of $\Delta_6$ to \ddbar
mixing observables and place constraints on its Yukawa coupling to $c$ and
$t$ quarks---$g_6^{23}$.

\begin{figure}[h]
\centering\begin{tabular}{lcr}
\includegraphics[width=0.24\textwidth]{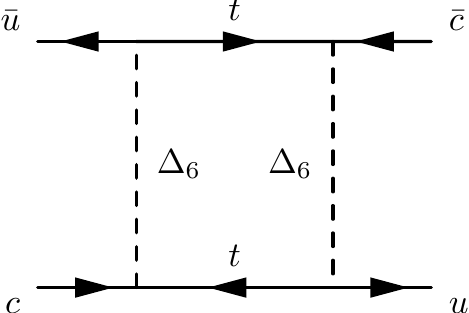}&&
\includegraphics[width=0.24\textwidth]{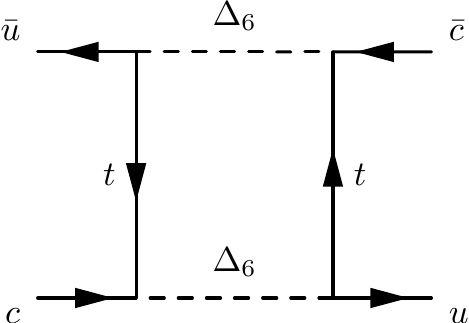}
\end{tabular}
\caption{Contribution of order $(g_6^{13} g_6^{23*})^2$ to $\Delta C =
  2$ effective Hamiltonian.}
\label{fig:DDbar}
\end{figure}

The \ddbar mixing matrix element of $|\Delta C|=2$ Hamiltonian is
split into dispersive ($M_{12}$) and absorptive ($\Gamma_{12}$) parts
as following
\begin{equation}
  \Braket{D^0 | \mc{H} | \bar D^0} = M_{12} - \frac{i}{2} \Gamma_{12},
\end{equation}
and due to the (anti)hermiticity of $M_{ij}$~($i/2\,\Gamma_{ij}$) also
$\Braket{\bar D^0 | \mc{H} | D^0} = M^*_{12} - \frac{i}{2}
\Gamma^*_{12}$ holds. Two dimensionless parameters describe
magnitudes of the dispersive~($x_{12}$) and the absorptive~($y_{12}$)
part while a relative phase between $M_{12}$ and $\Gamma_{12}$ would
signal $CP$ violation~\cite{Gedalia:2009kh}
\begin{align}
x_{12} &= \frac{2|M_{12}|}{\Gamma},\qquad y_{12} = \frac{|\Gamma_{12}|}{\Gamma},\qquad \phi_{12} = \arg(M_{12}/\Gamma_{12}). \label{eq:ThPars}
\end{align}
Here $\Gamma$ is the average decay width of neutral $D$
mesons. Contributions to the absorptive part $y_{12}$ come from
on-shell intermediate states and are thus accounted for by SM. On the
other hand, in the presence of $\Delta_6$ $x_{12}$ and $\phi_{12}$ are
affected by the box diagrams consisting of $\Delta_6$ and $t$-quark
exchanges which mediate \ddbar transitions, as shown on
Fig.~\ref{fig:DDbar}.
Since the preferred mass of $\Delta_6$ is between $300\e{GeV}$ and $1\e{TeV}$
in order to explain the $t\bar t$ asymmetry~\cite{Dorsner:2009mq},
we can safely integrate out both the top quark and $\Delta_6$ at a
common scale $\mu = m_{\Delta_6}$. Leading order matching onto the
effective theory generates a single operator, denoted in the
literature as $Q_6$~\cite{Ciuchini:1997bw}
\begin{equation}
  \label{eq:effham}
  \mc{H}(\mu = m_{\Delta_6}) = C_6(m_{\Delta_6}) Q_6, \qquad Q_6 = (\bar u_R \gamma^\mu c_R) (\bar u_R \gamma_\mu c_R),
\end{equation}
with the corresponding Wilson coefficient
\begin{equation}
  C_6(m_{\Delta_6}) = \frac{(g^{13}_6 g_{6}^{23*})^2 h(m^2_{\Delta_6}/m^2_t)}{32 \pi^2 m_t^2}, \qquad   h(x) = \frac{x^2-2x\log x-1}{(x-1)^3}.
\end{equation}
Effective Hamiltonian is evolved down to the charm scale $\mu_D=2\e{GeV}$ using
the leading-log anomalous dimension. The multiplicative
renormalization factor of $C_6$ in our case is adopted from
Ref.~\cite{Golowich:2009ii} and reads
\begin{equation}
  r(\mu_D,m_{\Delta_6}) = \left(\frac{\alpha^{(5)}_S(m_{\Delta_6})}{\alpha^{(5)}_S(m_b)}\right)^{6/23} \left(\frac{\alpha^{(4)}_S(m_b)}{\alpha^{(4)}_S(\mu_D)}\right)^{6/25}.
\end{equation}
The nonperturbative bag parameter $B_D$ corrects
the vacuum insertion approximation value of the mixing matrix element
\begin{equation}
  \label{eq:bag}
  \braket{D^0 | (\bar u_R \gamma^\mu c_R)\,(\bar u_R \gamma_\mu c_R) |\bar D^0} = \frac{2}{3} m_D^2 f_D^2 B_D.
\end{equation}
We use the value $B_D(\mu_D=2\e{GeV}) = 0.785$ calculated on the
lattice employing quenched Wilson fermions~\cite{Gupta:1996yt}. For
the decay constant we use the CLEO measured value $f_D =
0.206\e{GeV}$~\cite{:2008sq}. 

\begin{figure}[h]
  \centering\includegraphics[height=0.18\textheight]{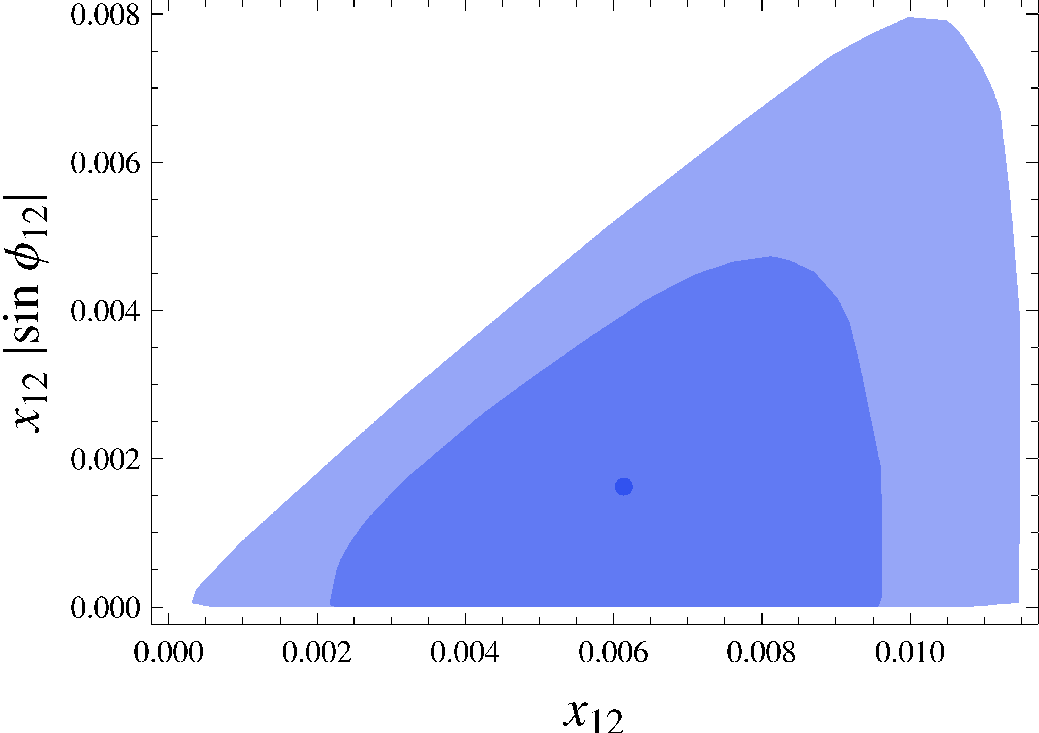}
  \caption{The $1$ and $2\sigma$ confidence level regions (dark and
    light-shaded regions, respectively) in the plane $x_{12}$--$x_{12}
    |\sin\phi_{12}|$.}
  \label{fig:x12phi12}
\end{figure}

\begin{figure}[h]
  \centering\includegraphics[height=0.4\textwidth]{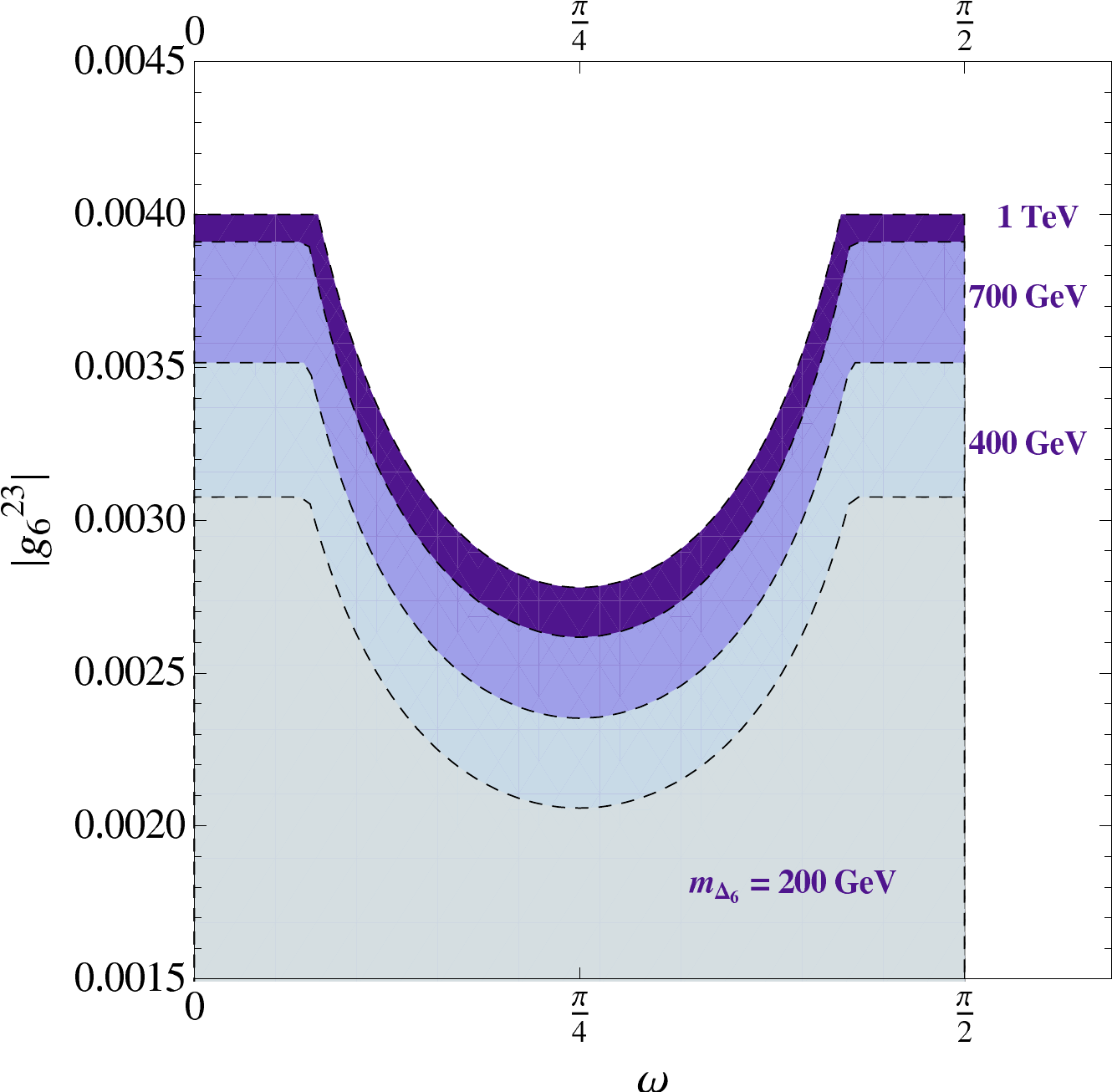}
  \caption{Allowed regions of $|g^{23}_6|$ and $\omega$ for different values of
    $m_{\Delta_6}$.}
  \label{fig:g23}
\end{figure}

Experimentally, \ddbar mixing has now been confirmed, while $CP$
violation is still consistent with zero, as is evident from the latest
HFAG~\cite{Barberio:2008fa} average which assumes no direct $CP$
violation
\begin{subequations}
\label{eq:DDbarmixing}
\begin{align}
&x = (0.59 \pm 0.20)\%, \qquad y = (0.81 \pm 0.13)\%,\\
&|q/p| = 0.98^{+0.15}_{-0.14}, \qquad \phi = -0.051^{+0.112}_{-0.115}\,.
\end{align}
\end{subequations}
The assumption of negligible direct $CP$ violation resides on the fact
that $\Delta_6$ cannot contribute to $\Gamma_{12}$, and that in the SM
$CP$ violation is very small~\cite{Gedalia:2009kh}. In this case, the
four measured parameters are governed by the three independent
theoretical quantities~(\ref{eq:ThPars}), implying redundancy of one
of the four experimental parameters. Indeed the values of $|q/p|$ and
$\phi$ are found to be strongly correlated in the HFAG
fit~(\ref{eq:DDbarmixing}) with correlation coefficient of
$0.614$~\cite{Barberio:2008fa}. We choose to extract $x_{12}$ and
$\phi_{12}$ from $x$, $y$, and $|q/p|$ using the following
relations~\cite{Gedalia:2009kh,Grossman:2009mn}
\begin{align}
  x_{12}^2 &= \frac{\left(|q/p|^2+1\right)^2 x^2 +
    \left(1-|q/p|^2\right)^2 y^2}{4
    |q/p|^2},\\
  \sin^2\phi_{12} &= \frac{\left(1-|q/p|^4\right)^2
    \left(x^2+y^2\right)^2}{16 |q/p|^4 x^2
    y^2+\left(1-|q/p|^4\right)^2 \left(x^2+y^2\right)^2}.
\end{align}
Imaginary part of $M_{12}$ is directly accessible in the product
\begin{equation}
  x_{12} \sin \phi_{12} = \frac{2\, \mrm{Im} M_{12}}{\Gamma},
\end{equation}
which we constrain, along with $x_{12}$, from the HFAG
values~(\ref{eq:DDbarmixing}).  The $1$ and $2\sigma$ confidence
levels of $x_{12}$ and $x_{12} |\sin \phi_{12}|$ are shown on
Fig.~\ref{fig:x12phi12},
whereas the upper bounds at $95$\,\% ($2\sigma$) confidence level
are
\begin{align}
  \label{eq:x12s12bounds}
  x_{12} < 9.6\E{-3},\qquad x_{12} |\sin \phi_{12}| < 4.4\E{-3}.
\end{align}
The imaginary part of $M_{12}$ originates from relative phase between
the $g_6^{23}$ and $g_6^{13}$, namely,
\begin{equation}
  \mrm{Im}\left[(g_6^{13} g_{6}^{23*})^2\right] = |g_6^{13}|^2 |g_6^{23}|^2 \sin (2\,\omega),
\end{equation}
where $\omega$ is the difference between phases of $g_6^{13}$ and
$g^{23}_6$. Using the central value for $|g_6^{13}|$, as given in
Eq.~\eqref{eq:g13fit}, we find the region of $|g_6^{23}|$, which is
limited from above by the $95$\,\% upper
bounds~(\ref{eq:x12s12bounds}), and is shown in Fig.~\ref{fig:g23}.
The bound on $x_{12} |\sin \phi_{12}|$ is stronger, except in the
regions close to $\omega=0$ or $\pi/2$, where bound on $x_{12}$
dominates. Apparent nondecoupling, i.e., strong bound even for large
masses $m_{\Delta_6}$, is due to linear dependence of $|g_6^{13}|$ on
$m_{\Delta_6}$ (see Eq.~(\ref{eq:g13fit})). We obtain a robust
$2\sigma$ bound in the region of interest $m_{\Delta_6} <
1\e{TeV}$:
\begin{equation}
  \label{eq:g23bound}
  |g^{23}_6| < 0.0038,
\end{equation}
regardless of complex phases, thus allowing also for finely tuned
phase $\omega$.

One can extract the product $|g_6^{13} g_6^{23*}|$ also from the radiative
$c \to u \gamma$ decay, where the $\Delta_6$-induced effective
Hamiltonian reads
\begin{equation}
  \label{eq:cTouGamma}
  \mc{H}^{c \to u \gamma} = \frac{g_6^{13} g_6^{23*} f(m_{\Delta_6}^2/m_t^2)}{6 m_t^2}\times \frac{e m_c}{(4\pi)^2} (\bar u_R \sigma^{\mu\nu} c_L) \, F_{\mu\nu},
\end{equation}
with $F_{\mu\nu}$, $e$ the electromagnetic tensor and coupling, whereas function $f$ is
\begin{equation}
  \label{eq:cugammaloop}
f(x) = \frac{2 x^3+3 x^2-6 x^2 \log x - 6x+1}{(x-1)^4}.
\end{equation}
However, using bounds on the couplings (Eqns.~\eqref{eq:g23bound} and
\eqref{eq:g13fit}) we find
 \begin{equation}
   \Gamma^{c \to u \gamma}/\Gamma_{D^0} \lesssim 10^{-10},
 \end{equation}
 whereas the SM resonant contributions to $D \to V \gamma$ and $D \to
 P \ell^+ \ell^-$ ($P$ and $V$ denote pseudoscalar and vector mesons)
 are of the order of
 $10^{-6}$~\cite{Fajfer:2007dy,Burdman:1995te,Burdman:2001tf}. This
 implies that radiative decays offer a far weaker constraint than the
 measured \ddbar observables.

\section{Bounds on $g_6^{12}$}
\label{g12}
\begin{figure}
 \includegraphics[height=.15\textheight]{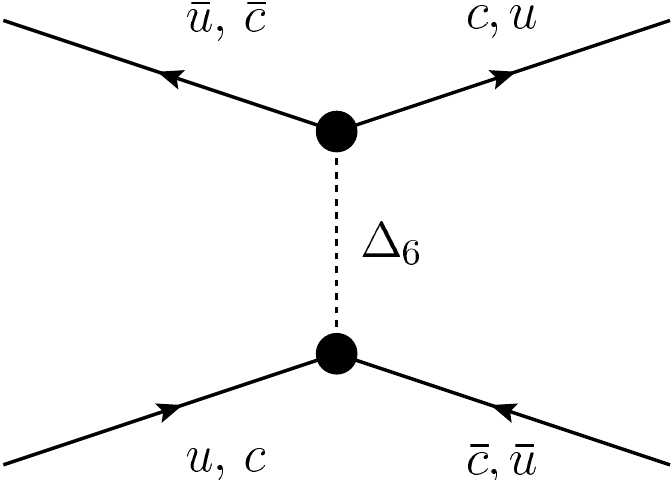}
  \caption{\label{fig:2jdiagrams} Partonic contribution to di-jet
    production at the Tevatron from a $u$-channel $\Delta_6$
    exchange. The analogous $s$- and $t$-channel contributions can simply
    be obtained via crossing.}
\end{figure}

\begin{figure}
  \vspace{1cm}
  \includegraphics[height=.22\textheight]{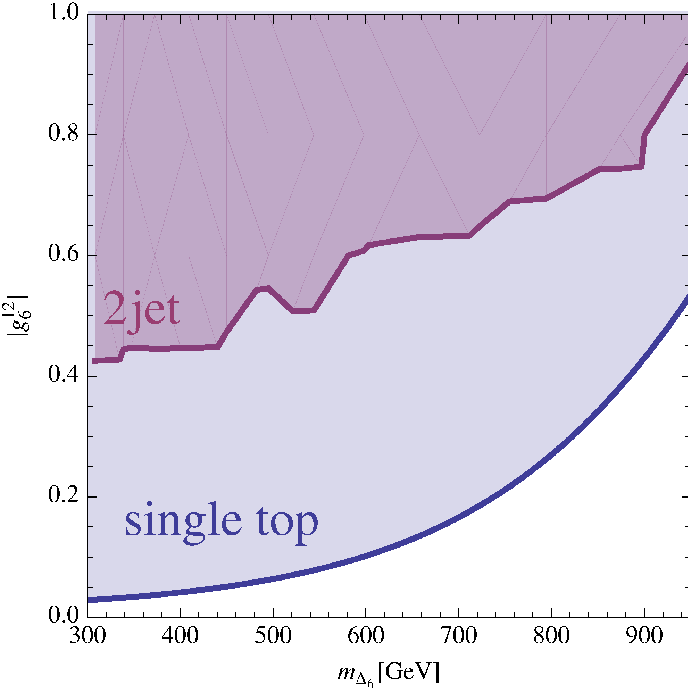}
  \caption{\label{fig:bounds} Constraint on the $|g_6^{12}|$ coupling
    and $\Delta_6$ mass from the single top production and di-jet at
    the Tevatron. The shaded areas are excluded.}
\end{figure}

In this section we consider bounds on the coupling $g_6^{12}$ coming
from the CDF search for resonances in the invariant mass spectrum of
di-jets~\cite{0812.4036} as well as from the single top production
cross-section measurements at the Tevatron~\cite{0908.2171}. The first
measurement constrains the $|g_6^{12}|$ coupling directly since the
process can be mediated by $\Delta_6$ through $s$-, $u$- and $t$-channel
exchange diagrams (see Fig.~\ref{fig:2jdiagrams}) interfering with leading order~(LO)
QCD contributions at the partonic level.
The resulting partonic $u\bar u \to c \bar c$ differential
cross-section from the $u$-channel $\Delta_6$ contribution interfering
with the $s$-channel single gluon exchange is
\begin{eqnarray}
  \frac{d\sigma^{u\bar u\to c\bar c}_6(\hat s)}{d\hat t}  &=& \frac{d\sigma_{SM}^{u\bar u\to c\bar c}(\hat s)}{d \hat t} 
  + \frac{{|g^{12}_6 |}^4 }{48 \pi \hat s^2  } \frac{ \hat u^2}{\left(m_{\Delta_6} ^2-\hat u\right)^2+\Gamma_{\Delta_6}^2} 
  \nonumber\\
  &&
  - \frac{  \alpha_s {|g^{12}_6|^2}}{9  \hat s^3}  \frac{\hat u^2\left( m_{\Delta_6}^2-\hat u \right)}{ \left(m_{\Delta_6} ^2-\hat u\right)^2 + \Gamma_{\Delta_6}^2}  
  \,,
\label{eq:sigmaLQ}
\end{eqnarray} 
where $\hat s=(p_{\bar u}+p_u)^2$, $\hat t=(p_{u}-p_c)^2$, $\hat
u=(p_{\bar u}-p_c)^2$. The expression for the $c\bar c \to u\bar u$ is
identical, while contributions from the related $u \bar c \to u \bar
c$, $c\bar u \to c \bar u$, $u c \to u c$ and $\bar u \bar c \to \bar
u \bar c$ processes can be obtained via crossing. Since the last two
contributions involve $\Delta_6$ exchange in the $s$-channel, we have
included the $\Delta_6$ total width $\Gamma_{\Delta_6}$ as the
regulator of the on-shell pole in all expressions. We assume that
$\Gamma_{\Delta_6}$ is dominated by the three decay channels to pairs
of quarks $ut$, $ct$ and $uc$. For the $\Delta_6 \to t q_i$ ($q_1,q_2=u,c$) channels we keep the top quark mass dependence in the partial width
 \begin{equation}
\Gamma(\Delta_6 \to t q_i) = \frac{|{g^{i 3}_{6}}|^2 \left(m_{\Delta_6}^2- m_t^2\right)^2}{16 \pi  m_{\Delta_6}^3}\,,
 \end{equation}
 while we neglect quark masses in the analogous expression for the
 $\Delta_6 \to u c$ channel. As shown in the previous section, the $t
 c$ channel is severely constrained by experimental results on \ddbar
 oscillations and can be neglected.  In order to compare with the
 experimental results~\cite{0812.4036} we first compute the hadronic
 di-jet production invariant mass spectrum by convoluting the LO QCD
 partonic differential cross-section including the above described
 tree-level $\Delta_6$ contributions with the CTEQ5 \cite{Lai:1999wy}
 set of parton distribution functions (PDFs) at the factorization and
 $\alpha_s$ renormalization scale set to $\mu_F=\mu_R = E_T$, where
 $E_T$ is the transverse energy of the final state di-jet. (We employ
 $\alpha_s(m_Z)=0.117$~\cite{Amsler:2008zzb}.) We also require both
 jets to have rapidity $|y|<1$. Then we reweigh our results so that
 our SM predictions (with $g_6^{12}=0$) match the complete NLO QCD
 results computed with FastNLO~\cite{FastNLO}, CTEQ6.1~\cite{CTEQ61}
 set of PDFs and appropriate jet algorithms as used in the original
 CDF analysis~\cite{0812.4036}. Finally, we turn on $\Delta_6$
 contributions and compare the resulting spectra with published CDF
 results. Assuming that $\Delta_6$ is responsible for the measured
 large FBA in top quark pair production at the
 Tevatron~\cite{FBAexp1,FBAexp2,FBAexp3}, combined with the measured
 total $t\bar t$ production cross-section as well as the top pair
 invariant mass spectrum, leads to the $g_6^{13}$ coupling and
 $\Delta_6$ mass satisfying the approximate relation given in
 Eq.~\eqref{eq:g13fit}. Then the total $\Delta_6$ decay width in the
 interesting region of $|g_6^{12}|$ is comparable with the
 experimental di-jet invariant mass bin size. We obtain the bounds on
 $|g_6^{12}|$ as a function of $m_{\Delta_6}$ by comparing the
 obtained theoretical spectrum for given values of $\Delta_6$
 parameters with the experimental measurement.  The results are shown
 as the purple shaded area in Fig.~\ref{fig:bounds}.

\begin{figure}
  \includegraphics[height=.15\textheight]{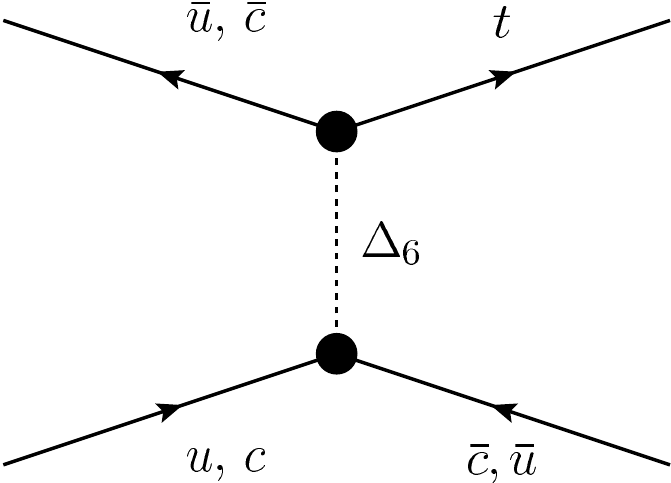}
  \caption{\label{fig:1topdiagrams} Partonic contribution to the
    single top production at the Tevatron from a $u$-channel $\Delta_6$
    exchange. The analogous $s$-channel contribution can simply be
    obtained by crossing.}
\end{figure}

 The single top production cross-section is sensitive to the product
 of $|g_6^{12} g_6^{13*}|$ (and also $|g_6^{12} g_6^{23*}|$, which,
 however, is more severely constrained by \ddbar oscillation
 measurements and we neglect it in what follows) since it proceeds at
 partonic level through $s$- and $u$-channel $\Delta_6$ exchange diagrams
 (see Fig.~\ref{fig:1topdiagrams}).
The resulting $u$-channel partonic contribution to $u\bar u \to t \bar
c$ has the form
\begin{equation}
\label{eq:st}
  \frac{d\sigma^{u\bar u \to t \bar c}}{d\hat t} = - \frac{|g_6^{13*} g_6^{12}|^2}{48 \pi \hat s^2} \frac {(\hat s + \hat t) \hat u}{(\hat u-m_{\Delta_6}^2)^2 + \Gamma_{\Delta_6}^2 }\,,
\end{equation}
where now $\hat s=(p_{\bar u}+p_u)^2$, $\hat t=(p_{u}-p_t)^2$, $\hat
u=(p_{\bar u}-p_t)^2$. 
The analogous $u c \to t u$
expression is related to it via crossing. A bound on $|g_6^{12}|$ can
then be obtained by using the $t\bar t$ FBA preferred values of
$|g_6^{13}|$ in Eq.~(\ref{eq:g13fit}). Since all the current
experimental measurements of single top production use sophisticated
multivariate analysis techniques it is difficult to compare $\Delta_6$
contributions directly with their data. Therefore we employ a
conservative approach and only compare NP contributions with the
experimental error on the combined Tevatron result for the total
single-top production cross-section (summed over both $t$ and $\bar
t$) of $\sigma_{1t} = 2.76^{+0.58}_{-0.47}$~pb~\cite{0908.2171}---we
require the $\Delta_6$ mediated hadronic production cross-section,
which we denote $\Delta\sigma_{1t}$, to be smaller than $1$~pb at
$95\%$ confidence level. A recent dedicated D0 analysis of NP mediated
single top production obtained even stricter limits on the anomalous
single top production cross-section~\cite{1006.3575}, however their
results are not directly applicable to our case due to different
partonic initial and final states as well as different kinematics. We
compute the NP-mediated single top production by again convoluting the
partonic differential cross-section with the appropriate PDFs at the
factorization and $\alpha_s$ renormalization scale set to $\mu_F=\mu_R
= E_T$ and then integrating over the available hadronic
phase-space. Since in this case there are no contributions due to the
SM-NP interference, the $\Delta_6$ mediated single top production
cross-section (and the associated constraint) scales quadratically
with $|g_6^{12} g_6^{13*}|$. Using the $t\bar t $ forward backward
asymmetry preferred values of $|g_6^{13}|$ in Eq.~\eqref{eq:g13fit} we
obtain the bounds on $|g_6^{12}|$ shaded in blue in
Fig.~\ref{fig:bounds}.

\section{Predictions for rare top quark decays}
\label{topdecays}
Having obtained the upper bounds on $|g_6^{12}|$ and $|g_6^{23}|$ we
can now assess the prospects of observing radiative top quark decays
at the LHC. We predict decay widths of processes with an up-type quark
and a photon or a gluon in the final state.

The $\Delta_6$-induced effective Hamiltonian for a top quark decay to a
light up-type quark (massless $c$ or $u$) and a single photon or a gluon reads
\begin{align}
  \label{eq:tTocGamma}
  \mc{H}^{t \to q_i \gamma, q_i G} =& g_6^{12} \frac{F^\gamma(m_{\Delta_6}^2/m_t^2)}{3 m_t^2}\,\, \frac{e m_t}{(4\pi)^2}
  \left[g_6^{13*} (\bar c_R\sigma^{\mu\nu} t_L) - g_6^{23*} (\bar u_R
    \sigma^{\mu\nu} t_L)\right] \, F_{\mu\nu}\\
  -&g_6^{12}\frac{F^G(m_{\Delta_6}^2/m_t^2)}{2 m_t^2}\,\, \frac{g
    m_t}{(4\pi)^2} \left[g_6^{13*} (\bar c_{R}\sigma_{\mu\nu}
    T^A t_{L})-g_6^{23*} (\bar
    u_{R}\sigma_{\mu\nu} T^A t_{L})\right]
  G_A^{\mu\nu},\nonumber
\end{align}
where $e$, $F_{\mu\nu}$ and $g$, $G^A_{\mu\nu}$ are electromagnetic
and color coupling constant and field tensors, respectively. Dependence
on $m_{\Delta_6}$ is ascribed to
\begin{subequations}
\begin{align}
F^\gamma(x) &= -2 x - 3 -2 (x-1) x \log \frac{x-1}{x} +4 x\, \mrm{Li}_2(1/x),\\
F^G(x) &= 2 x + 2x (x-1)\log\frac{x-1}{x} -x\, \mrm{Li}_2(1/x).
\end{align}
\end{subequations}
The light quark in the final state, regardless of being either $c$ or
$u$, couples via $g_6^{12}$ to the light quark and $\Delta_6$ in the
loop.  The top quark, however, couples predominantly to $u$ and
$\Delta_6$ (its coupling to $c$ and $\Delta_6$ is suppressed by the smallness of
$|g_6^{23}|$) in the loop and thus $t \to u \gamma, u G$ decay widths
are suppressed by a factor of $|g_6^{23}/g_6^{13}|^2 \sim 10^{-6}$ with
respect to $t \to c \gamma, c G$ widths. For decays with a $c$-quark
in the final state we find at leading order in $\alpha_S$
\begin{equation}
\label{eq:tWidths}
  \Gamma^{t \to c \gamma} = \frac{\alpha |g_6^{12} g_6^{13}|^2 m_t}{2304 \pi^4} \left[F^\gamma(m^2_{\Delta_6}/m_t^2)\right]^2,\qquad \Gamma^{t \to c G} = \frac{\alpha_S |g_6^{12} g_6^{13}|^2 m_t}{768 \pi^4} \left[F^G(m^2_{\Delta_6}/m_t^2)\right]^2.
\end{equation}
For $\Delta_6$ masses below $1\e{TeV}$ the
expressions~\eqref{eq:tWidths} lead to branching fractions' upper
limit of the order of $10^{-9}$ for both $t \to c \gamma$ and $t \to c
G$. Independently of the coupling constant values, the width of the
photonic channel is $30$--$40$\,\% larger than the $t\to c G$ width.
QCD corrections for these processes are
known~\cite{Zhang:2008yn,Drobnak:2010wh,Zhang:2010bm}. For the $t\to
cG$ channel they amount to 20\,\% enhancement in the branching
fraction. On the other hand, the smallness of the gluonic rate leads
to negligible effects in the $t\to c\gamma$ branching fraction~\cite{Drobnak:2010wh}.

\begin{figure}[h]
  \includegraphics[height=.22\textheight]{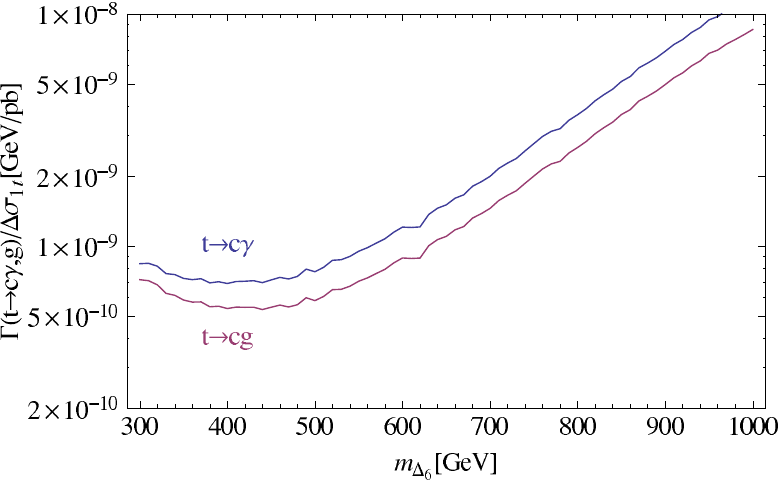}
  \caption{Ratio of top quark decay widths $t \to c \gamma, c G$ and
    $\Delta_6$ contribution to the single top production at the
    Tevatron. See text for explanation.}
  \label{fig:corr}
\end{figure}

Interestingly, the ratio of tree-level contributions of $\Delta_6$ in
single $t$ production cross-section $\Delta\sigma_{1t}$ (see
Eq.\eqref{eq:st}) and $t \to c \gamma$ or $t \to c G$ decay width is
completely independent of $g_6$ couplings. On the other hand, the
ratio $\Gamma_{t \to c G, c\gamma}/\Delta\sigma_{1t}$ exhibits a
nontrivial $\Delta_6$ mass dependence, as shown in
Fig.~\ref{fig:corr}.

\section{Prediction for the up-quark Yukawa sector}
\label{upyukawa}
The fact that all three entries of $g_6$ are either fixed or bounded
through processes that involve the same scalar field, i.e.,
$\Delta_6$, presents us with a unique opportunity to put constraints
on relevant Yukawa coupling constants within a large class of GUT
models in the up-quark sector at the very high scale as we show next.

We mentioned already that $\Delta_6$ is a part of a $45$-dimensional Higgs representation
of $SU(5)$ and the $45$-dimensional Higgs representation, i.e.,
$\bm{45}$, is frequently used in $SU(5)$ GUT models. The
relevant coupling---already given in Eq.~\eqref{M_D&M_E}---basically
sets a model dependent lower bound on a vacuum expectation value of
the Higgs doublet that resides in $\bm{45}$.

We find that at least one component of $g_6$ that describes the
couplings of $\Delta_6$ to the up-quarks must differ from zero to
accommodate forward-backward asymmetry in top quark pair production at
the Fermilab Tevatron. But, $g_6$ is related, via Eq.~\eqref{g6}, to
the coupling of the Higgs doublet in the $45$-dimensional
representation to matter. This antisymmetric contribution, however, is
not sufficient to generate viable up-quark mass spectrum. Namely, one
$5$-dimensional Higgs, i.e., $\bm{5}$, is also needed to generate
viable up-quark mass spectrum via renormalizable set of $SU(5)$ Yukawa
coupling contractions. The relevant contraction of $\bm{5}$ with
matter is $\epsilon_{\alpha \beta \gamma \delta \epsilon} (Y'_2)_{ij}
(\bm{10}^{\alpha \beta})_i (\bm{10}^{\gamma \delta })_j
(\bm{5})^{\epsilon}$. The up-quark mass matrix at the GUT scale reads
  \begin{equation}
  \label{M_U}
M_U = \left[ 4 (Y'^T_2+Y'_2) v_5  -  8  (Y^T_2-Y_2) v_{45}\right]/\sqrt{2},
\end{equation}
  where $\langle\bm{5}\rangle^5=\sqrt{2} v_5$ and
  $\langle\bm{45}\rangle^{5 1}_{1}= \langle\bm{45}\rangle^{5 2}_{2}=\langle\bm{45}\rangle^{5 3}_{3} =\sqrt{2} v_{45}$ represent
appropriate vacuum expectation values.
   (Note, $2 |v_5|^2+48 |v_{45}|^2=v^2$, where $v=246$\,GeV.) 
 $Y'_2$ and $Y_2$ are in general arbitrary $3 \times 3$ Yukawa matrices.

  We can now show that constraints on $g_6$ immediately translate into
  constraints on symmetric coupling contribution in
  Eq.~\eqref{M_U}. Namely, if we go to the basis where the down-quarks
  mass matrix is diagonal we obtain the following two equations
\begin{subequations}
\label{S&A}
\begin{align}
&4 S' = U^\dag M_U^{diag}+ M_U^{diag} U^*\\
&4 A' = U^\dag M_U^{diag}- M_U^{diag} U^*,
\end{align}
\end{subequations}
where $U=\tilde V_{CKM} U_R$ and $M_U^{diag}$ is a diagonal up-quark mass
matrix. Note that $\tilde V_{CKM}=U_1 V_{CKM} U_2$ is proportional to
Cabibbo-Kobayashi-Maskawa (CKM) matrix $V_{CKM}$ apart from five phases
that are present at the GUT scale contained within diagonal unitary
matrices $U_1$ and $U_2$.
$A'(=2 \sqrt{2} U_R^\dag
(Y_2-Y^T_2) U_R^* v_{45})$ and $S'(=\sqrt{2} U_R^\dag (Y'^T_2+Y'_2)
U_R^* v_{5})$ are antisymmetric and symmetric contributions,
respectively. $U_R$, again, is an arbitrary unitary matrix. These
relations hold at the GUT scale whereas our constraints on $g_6$
originate from the low-energy phenomenology. If we run $g_6$
constraints as well as relevant fermion masses and their mixing parameters
from
electroweak scale to the GUT scale we would finally have $A'=g_6
v_{45}$. It is this relation that lets us deduce generic properties of
the symmetric contribution $S'$ and thus pinpoint the texture of the
up-quark Yukawa couplings.

We now state several important generic features of the Yukawa coupling
constants in the up-quark sector as implied by our constraints on
$g_6$. Firstly, the ratio $|S'_{13}|/|A'_{13}|$ is equal to one at all
instances. This is due to the fact that one of the $|A'|$ entries,
i.e., the one proportional to $|g_6^{13}|$ is fixed phenomenologically
and is significantly larger than the upper bounds on the other two
entries of $|A'|$. Further, both $|A'_{13}|$ and $|S'_{13}|$ are well
approximated with the same product $|U_{31}| m_t$ since $M_U^{diag}$
is highly hierarchical matrix. Since $|g_6^{13}|$ is fixed and larger
than both $|g_6^{12}|$ and $|g_6^{23}|$, we find $|U_{31}|/v_{45}$ to be
just a constant for a given $m_{\Delta_6}$ while $|S'_{13}|=|A'_{13}|$.

Secondly, since $|A'_{23}|$ should phenomenologically be much
smaller when compared with $|A'_{13}|$ it is clear that $|U_{32}|$
entry must be very small in order to suppress the first term in
$A'_{23}=U^*_{32} m_t-U^*_{23} m_c$ equality. As a consequence, there
is thus a correlation between $|U_{31}|$ and $|U_{33}|$ due to
unitarity of $U$. This correlation also implies that the ratio
$|S'_{33}|/|A'_{13}|$ should depend only on $|U_{31}|$ (or,
interchangeably, $v_{45}$). In fact, analytically we find
$|S'_{33}|/|A'_{13}| \approx 2 \sqrt{1-|U_{31}|^2}/|U_{31}|$. We plot
$|U_{31}|$ vs.\ $|S'_{33}|/|A'_{13}|$ as obtained in our numerical
analysis superimposed over $2 \sqrt{1-|U_{31}|^2}/|U_{31}|$ in
Fig.~\ref{figure:THREE}. Squares represent satisfactory numerical
solutions of Eqns.~\eqref{S&A} that satisfy all other phenomenological
constraints. We describe our numerical procedure in detail later
on. The most important facts we can conclude from
Fig.~\ref{figure:THREE} are that $|U_{32}|$ is indeed negligible,
$|S'_{33}|$ is determined for a given value of $|U_{31}|$ and
$|S'_{33}|$ is usually smaller than or, at most,
comparable with $|A'_{13}|$.

Thirdly, there is a connection between $|S'_{12}| =|U^*_{21} m_c+U^*_{12}
m_u|$ and $|A'_{12}|=|U^*_{21} m_c-U^*_{12} m_u|$. Namely, if
$|U_{31}|$ is small then unitarity makes $|U_{21}|$ large which in
turn implies that the ratio $|S'_{12}|/|A'_{12}|$ tends to one. In the
other extreme, when $|U_{31}|$ approaches one, $|U_{21}|$ goes to zero
and there is a possible interference between the two terms in
$|S'_{12}|$ and $|A'_{12}|$ which makes their ratio deviate from one
but not by a large amount. This behavior is depicted in
Fig.~\ref{figure:ONE} where we plot $\log |S'_{12}|/|A'_{12}|$ vs.\
$|U_{31}|$. Clearly, $|S'_{12}|$ and $|A'_{12}|$ are within a factor of
three from each other for all values of $|U_{31}|$. Note, $|U_{31}|$
cannot be arbitrarily small as is also
evident in Figs.~\ref{figure:THREE} and~\ref{figure:ONE}. In fact, when
$|U_{31}| \rightarrow 0$, with $|U_{32}|$ negligible, $|A'_{13}|
\rightarrow 0$.

Fourthly, there exist a correlation that concerns $|S'_{11}|$ and
$|S'_{22}|$. What we find is that, due to the fact that $|U_{32}|$ is
negligible,
unitarity of $U$ demands that $|S'_{11}|/|S'_{22}| \rightarrow m_u/m_c$
for $|U_{31}| \rightarrow 0$. We plot this behavior in
Fig.~\ref{figure:FOUR}, where we plot $\log |S'_{11}|/|S'_{22}|$ vs.\
$|U_{31}|$. It is clear from Fig.~\ref{figure:FOUR}
that $|U_{31}| \neq 1$ yields unique value for $|S'_{11}|/|S'_{22}|$.
Recall, in that regime, the ratio $|S'_{12}|/|A'_{12}|$ is also uniquely
determined.

Finally, there exists a model dependent lower bound (and upper bound)
on $v_{45}$ from the charged lepton and down-quark sector in models
with the 45-dimensional Higgs. However, due to the fact that the
$\bm{45}$ contraction on its own cannot provide viable up-quark masses
it is clear that there will also exist an upper bound on $v_{45}$ from
the observed up-quark sector masses. The novelty is that this upper
bound will be $m_{\Delta_6}$ dependent since the phenomenologically
viable form of $g_6$ is $m_{\Delta_6}$ dependent as given in
Eq.~\eqref{eq:g13fit}. We show the upper bound on $v_{45}$ as a
function of $m_{\Delta_6}$ in Fig.~\ref{figure:TWO}. As expected, it
drops as $m_{\Delta_6}$ grows in order to suppress corresponding
growth of $|g_6^{13}|$.

Before we turn to numerical analysis we stress that all our previous
observations do
not depend on exact values of up-quark masses.

\subsection{Numerical analysis}
\begin{figure}
\includegraphics[height=.34\textheight]{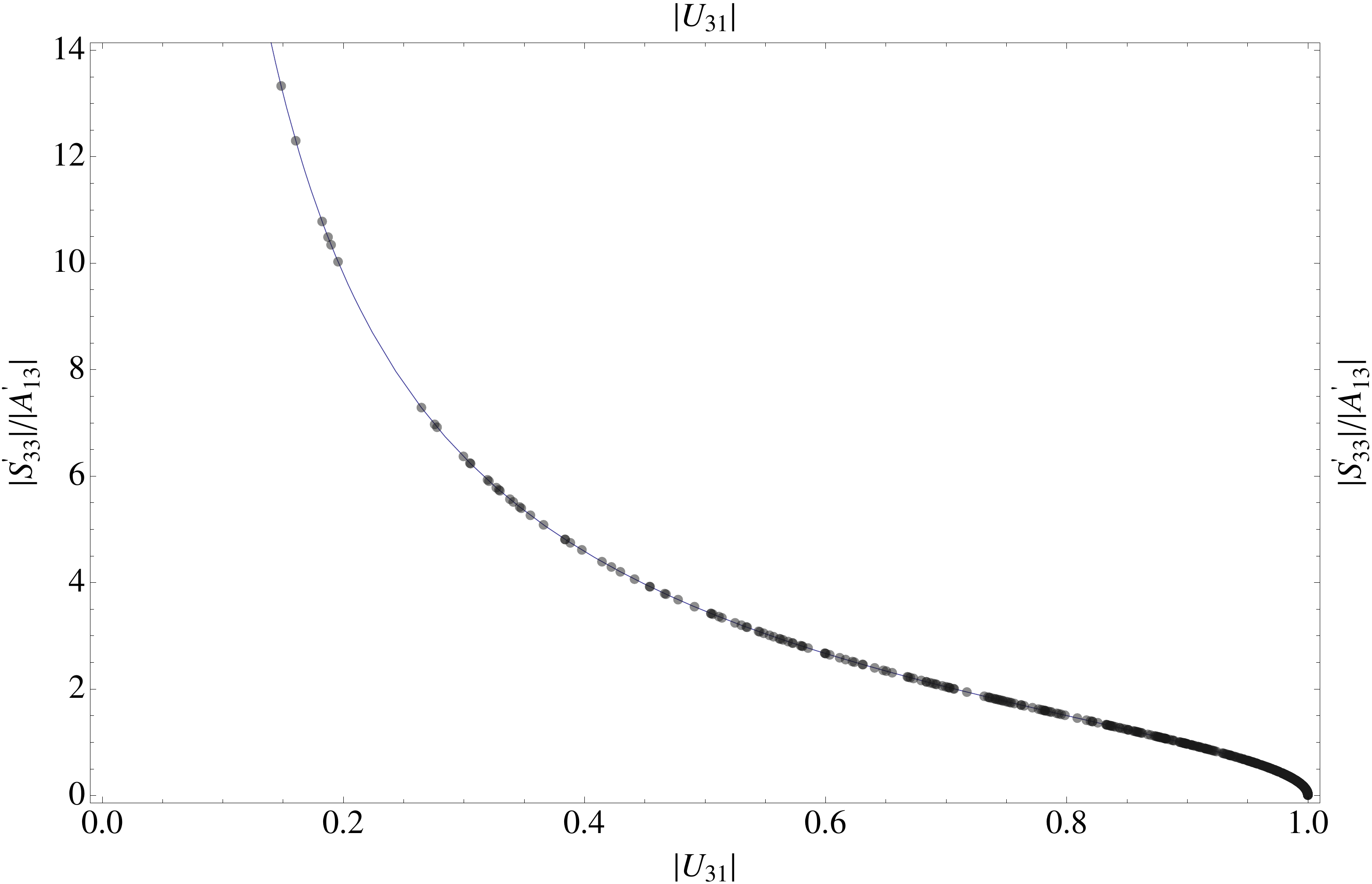}
\caption{\label{figure:THREE} Plot of $|U_{31}|$ vs.\
  $|S'_{33}|/|A'_{13}|$. Squares represent results of our numerical
  analysis as described in the text whereas the curve stands for
  approximate analytic expression $2
  \sqrt{1-|U_{31}|^2}/|U_{31}|$. The observed agreement implies that
  $|U_{32}|$ is negligible.}
\end{figure}

\begin{figure}
\includegraphics[height=.34\textheight]{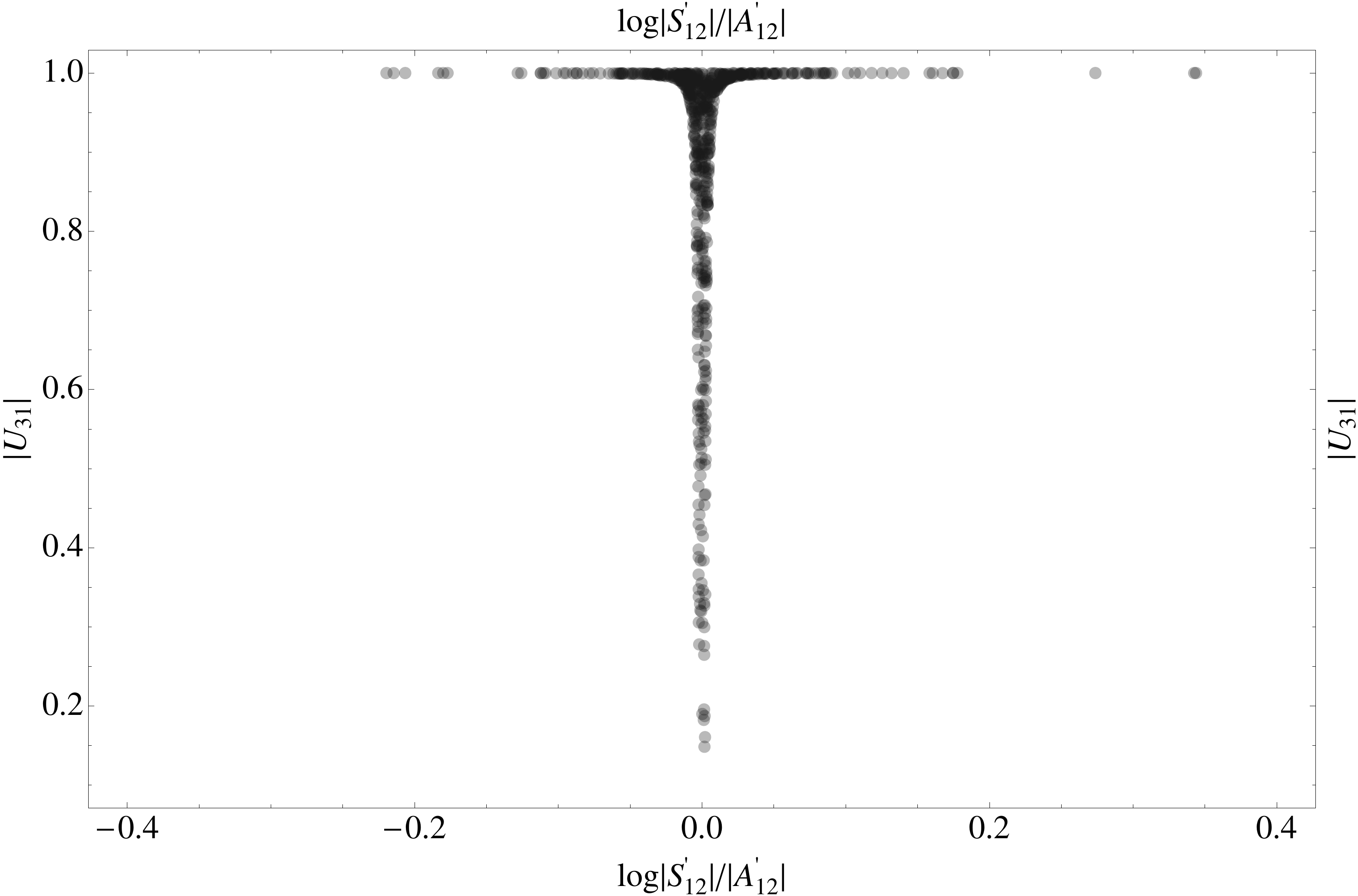}
\caption{\label{figure:ONE} Plot of $\log |S'_{12}|/|A'_{12}|$ vs.\
   $|U_{31}|$. There exist a lower bound on $|U_{31}|$ due to the fact
   that $|A'_{13}|>|A'_{23}|$. See text for details.}
\end{figure}

\begin{figure}
\includegraphics[height=.34\textheight]{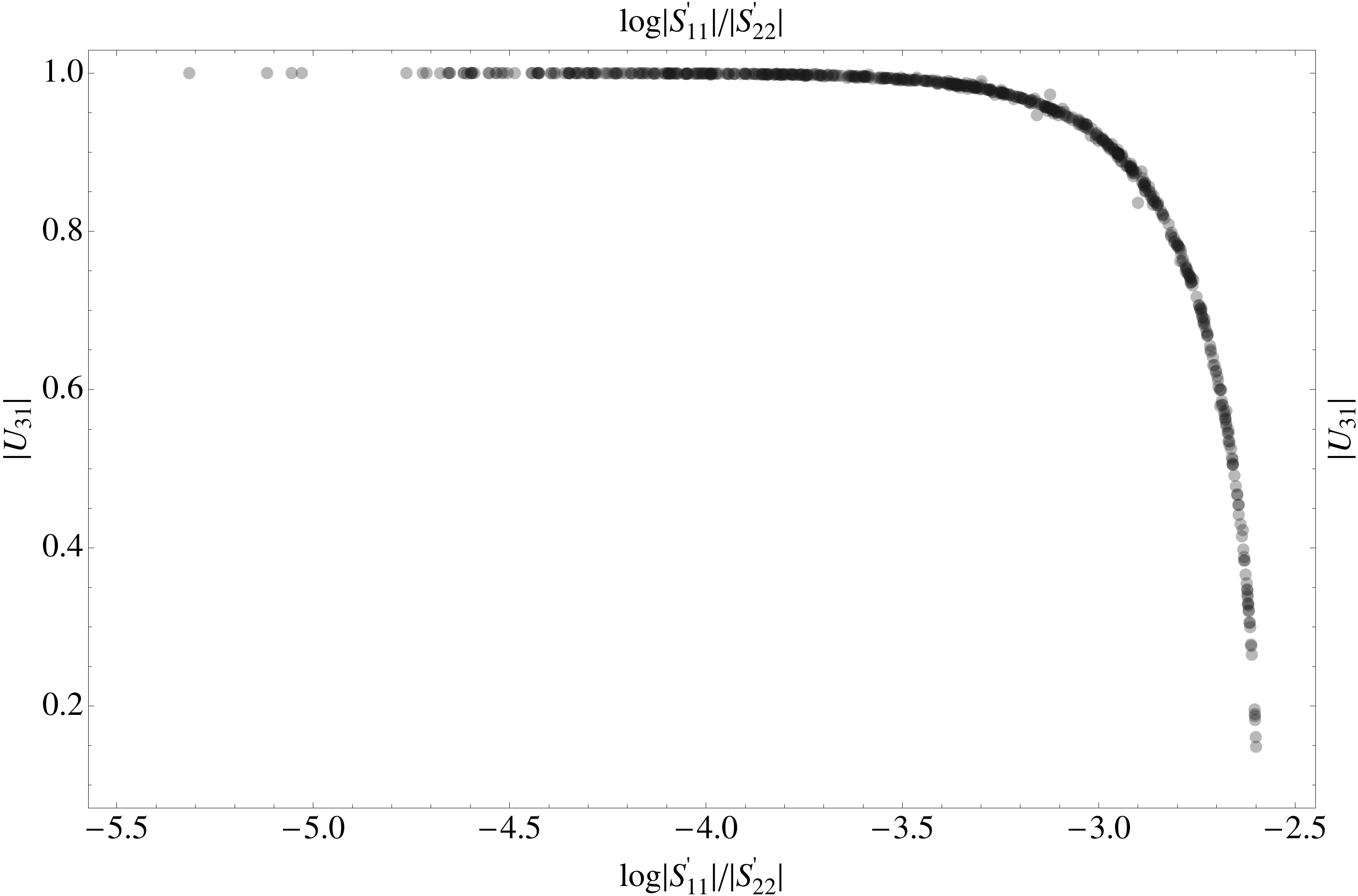}
\caption{\label{figure:FOUR} Plot of $\log |S'_{11}|/|S'_{22}|$ vs.\
   $|U_{31}|$. $|S'_{11}|/|S'_{22}| \rightarrow m_u/m_c$ for small
$|U_{31}|$.}
\end{figure}

\begin{figure}
\includegraphics[height=.34\textheight]{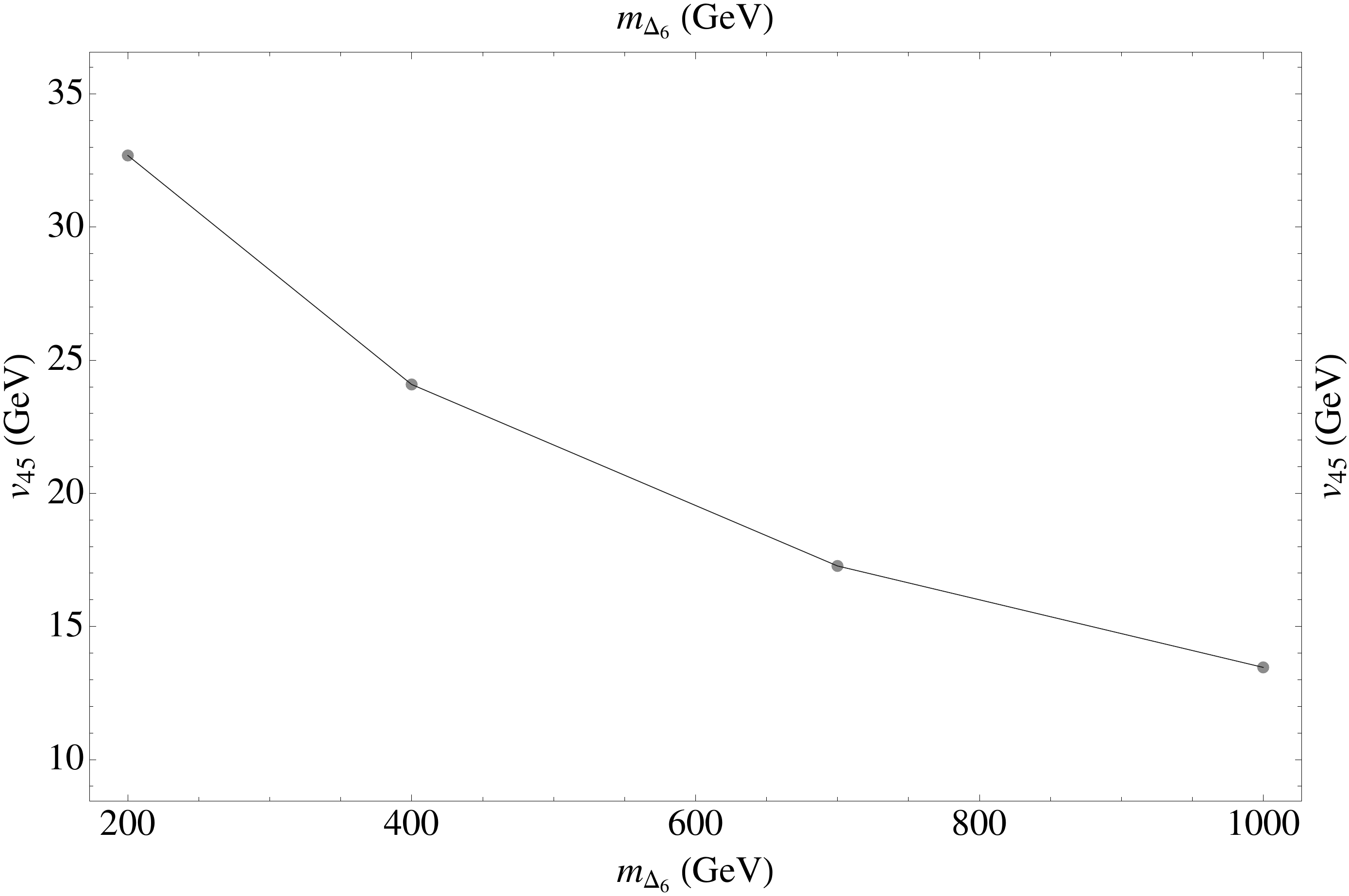}
\caption{\label{figure:TWO} The dependence of upper bound on $v_{45}$
   as a function of $m_{\Delta_6}$. Note that $v_{45}$ drops as
   $m_{\Delta_6}$ grows to decrease the influence of the antisymmetric
   components of $g_6$ towards the up-quark mass relations.}
\end{figure}

To generate our numerical results shown in
Figs.~\ref{figure:THREE},~\ref{figure:ONE},~\ref{figure:FOUR}
and~\ref{figure:TWO} we
have turned to a specific scenario comprising $\bm{45}\equiv(\Delta_1, \Delta_2, \Delta_3, \Delta_4, \Delta_5,
\Delta_6, \Delta_7) = (\bm{8},\bm{2},1/2)\oplus
(\overline{\bm{6}},\bm{1}, -1/3) \oplus (\bm{3},\bm{3},-1/3)
\oplus (\overline{\bm{3}}, \bm{2}, -7/6) \oplus (\bm{3},\bm{1},
-1/3) \oplus (\overline{\bm{3}}, \bm{1}, 4/3) \oplus (\bm{1},
\bm{2}, 1/2)$, $\bm{5}\equiv (\Psi_D,
\Psi_T) = (\bm{1},\bm{2},1/2)\oplus(\bm{3},\bm{1},-1/3)$ and
$\bm{24}\equiv(\Sigma_8
, \Sigma_3, \Sigma_{(3,2)}, \Sigma_{(\overline{3},2)},
\Sigma_{24}) = (\bm{8},\bm{1},0)\oplus(\bm{1},\bm{3},0)
\oplus(\bm{3},\bm{2},-5/6)\oplus(\overline{\bm{3}},\bm{2},5/6)
\oplus(\bm{1},\bm{1},0)$ of Higgs and one fermionic adjoint
representation~\cite{Perez:2007rm} $\bm{24}_F \equiv (\rho_8,\rho_3, \rho_{(3,2)}, \rho_{(\bar{3}, 2)},
\rho_{24})=(\bm{8},\bm{1},0)\oplus(\bm{1},\bm{3},0)\oplus(\bm{3},\bm{2},-5/6)
\oplus(\overline{\bm{3}},\bm{2},5/6)\oplus(\bm{1},\bm{1},0)$. Recall, $\Delta_6$ can be light in
this model without supersymmetry that employs $24$-dimensional
fermionic representation~\cite{Perez:2007rm} to generate neutrino
masses via combination of type
I~\cite{Minkowski:1977sc,Yanagida:1979as,GellMann:1980vs,Glashow:1979nm,Mohapatra:1979ia}
and type III~\cite{Foot:1988aq,Ma:1998dn} seesaw mechanisms. (This is
a renormalizable version of the model first proposed
in~\cite{Bajc:2006ia} and further analyzed
in~\cite{Dorsner:2006fx,Bajc:2007zf}.) In fact, we have demonstrated~\cite{Dorsner:2009mq}
that this scenario predicts proton decay signatures that are very
close to the present experimental limits due to partial proton decay
lifetime measurements when both $\Delta_1$ and $\Delta_6$ are in the
range accessible in collider experiments.

Numerical procedure we use to generate particle spectrum aims at an exact 1-loop level gauge coupling unification scenario with maximal possible unification scale $M_{GUT}$. It allows for multiple
particle thresholds as long as all relevant constraints are satisfied. In particular, we require that $10^2$\,GeV$\leq m_{\Sigma_3}, m_{\Sigma_8}, m_{\Delta_1}, m_{\Delta_2}, m_{\Delta_4}, m_{\Delta_7}, m_{\rho_3}, m_{\rho_{(3,2)}}, m_{\rho_{(\bar{3}, 2)}} \leq M_{GUT}$, $10^{12}$\,GeV$\leq m_{\Psi_T}, m_{\Delta_3}, m_{\Delta_5} \leq M_{GUT}$ and $10^6$\,GeV$\leq m_{\rho_8} \leq M_{GUT}$. Note, the masses of $\Psi_T$, $\Delta_3$ and $\Delta_5$ are bounded from below due to the fact that these fields mediate proton decay through the $d=6$ operators. $\rho_8$, on the other hand, should be heavier than $10^{6}$\,GeV to accommodate the Big Bang Nucleosynthesis constraints~\cite{Bajc:2006ia}. Finally, the allowed couplings of the fermionic adjoint lead to the following mass relations~\cite{Perez:2007rm}
\begin{equation}
m_{\rho_8}=\hat{m}m_{\rho_3}, \qquad m_{\rho_{(3,2)}}=m_{\rho_{(\bar{3}, 2)}}=\frac{(1+\hat{m})}{2}m_{\rho_3},
\end{equation}
where $\hat{m}$ represents a free parameter that describes the mass splitting between $\rho_8$ and $\rho_3$ states.

We take the relevant parameter for the adjoint fermion
mass spectrum to be $\hat{m} = 10^{14}$ and set $m_{\Delta_6}=400$\,GeV and $m_{\Delta_1}=1$\,TeV to
generate particle spectrum of the theory. This set of assumptions together with $\alpha_3 = 0.1176$, $\alpha^{-1} = 127.906$ and $\sin^2 \theta_W = 0.23122$~\cite{Amsler:2008zzb} yields $M_{GUT}=1.1 \times 10^{16}$\,GeV and $\alpha^{-1}_{GUT}=26.9$, where
$\alpha_{GUT}$ represents unified gauge coupling at $M_{GUT}$. The corresponding particle spectrum reads 
$m_{\Delta_2}=2.6$\,TeV, $m_{\Delta_3}=10^{12}$\,GeV, $m_{\Delta_4}= m_{\Delta_5}=1.1\times 10^{16}$\,GeV, $m_{\Delta_7}= 10^{2}$\,GeV, $m_{\Sigma_3}= 10^{2}$\,GeV, $m_{\Sigma_8}= 10^{2}$\,GeV, $m_{\Psi_T}=1.1 \times 10^{16}$\,GeV, $ m_{\rho_8} = 5.6 \times 10^{15}$\,GeV, $ m_{\rho_3} = 113$\,GeV and $m_{\rho_{(3,2)}}= m_{\rho_{(\bar{3}, 2)}} = 1.1 \times 10^{16}$\,GeV. Using that spectrum we find the 
following values of the up-quark masses and CKM parameters at the GUT scale at 1-loop level: $m_u =
0.00046$\,GeV, $m_c = 0.182$\,GeV, $m_t = 55.4$\,GeV, $s^{CKM}_{12}=0.225$, $s^{CKM}_{23}=0.00394$, $s^{CKM}_{13}=0.0462$ and $\delta^{CKM}=1.185$. The input values
at the $M_Z$ are $m_u = 0.0016$\,GeV, $m_c =
0.628$\,GeV, $m_t = 171.5$\,GeV, $s^{CKM}_{12}=0.2272$, $s^{CKM}_{23}=0.0422 $, $s^{CKM}_{13}=0.00399$ and $\delta^{CKM}=0.995$. All relevant input parameters, i.e., quark and lepton masses and mixing parameters, and running procedures are specified in
Ref.~\cite{Dorsner:2009mq}. 

In this particular regime proton decay contributions
are dominated by the color triplets that reside in both $\bm{45}$ and
$\bm{5}$. (See Fig.~[1] in Ref.~\cite{Dorsner:2009mq}.) However, to be
able to accurately predict partial decay
lifetimes we need to know the coupling strengths of the triplets
to the matter. Here, again, we demonstrate that one part---the one
related to the up-quark sector---is already well known due to
low-energy constraints. We have done the running of all relevant
parameters from low scale to high scale only for the
$m_{\Delta_6}=400$\,GeV case and used those values to infer that the
overall Yukawa coupling drop from low-scale to the GUT scale is well
described by a common factor of $3.5$. It is this factor that was used
for analysis for different values of $m_{\Delta_6}$. We find this
approximation well-justified since our main goal is to present generic
features of the up-quark Yukawa sector at the GUT scale.

We randomly vary angles and phases of $U$ to see if relevant
constraints on the elements of $A'$ are satisfied as given in
Eqns.~\eqref{eq:g13fit},~\eqref{eq:x12s12bounds} and in
Fig.~\ref{fig:bounds} adjusted by a factor of $3.5$. Once this is done
we read off phenomenologically allowed elements of $S'$. Our procedure
guaranties viable values of the up-quark masses and CKM mixing
parameters while $A'$ components satisfy all low energy
constraints. Note, since we know $V_{CKM}$ at the GUT scale, we can also
reproduce $U_R$ (up to some
phases of diagonal unitary matrices).

All in all, we have generated $10^8$ random points in the
nine-dimensional space of the unitary matrix $U$ for different values
of $\Delta_6$ mass, i.e., $m_{\Delta_6}=200, 400, 700, 1000$\,GeV. For
example, when $m_{\Delta_6}=400$\,GeV we find that out of $10^8$ sets
of initial values of parameters of $U$, i.e., the angles and phases of
unitary matrix, only 657 pass all phenomenological constraints which
implies rather unique form of Yukawa couplings. Due to the fact that
our antisymmetric matrix $A'$ is highly skewed with $|A'_{13}|$
element being dominant and the fact that the symmetric element
$|S'_{13}|$ is much larger than both $|S'_{12}|$ and $|S'_{23}|$
elements we have a situation that the underlying matrix $M_U$ as given
in Eq.~\eqref{M_U} in the down-quark mass eigenstate basis is of the
so-called lopsided form. (Recall, $|S'_{13}|$ or equivalently
$|A'_{13}|$ is either comparable or larger than $|S_{33}|$.)

Our results for the underlying Yukawa structure of the up-quark sector exhibit
dependence on $v_{45}$ (or $|U_{31}|=4 |g_6^{13}| v_{45}/m_t$). This, as we discuss, provides basis for analytic
correlation between all entries of $S'$ and $A'$, except for
$|S_{23}'|$. In order to appreciate the advantage of this analytic
study, we bracket the range of values that these entries can take by
resorting to our numerical analysis.  We present these ranges, without
correlation, for $m_{\Delta_6}=400$\,GeV and $m_{\Delta_6}=1$\,TeV in
Table~\ref{tab:table1} for the absolute values of components of
antisymmetric and symmetric contributions. Note that the ranges are
more or less $m_{\Delta_6}$ independent since $g_6^{ij}$ elements
scale similarly with respect to change in $m_{\Delta_6}$ in the region
of interest.
\begin{table}[h]
\caption{\label{tab:table1} Allowed ranges of absolute values of relevant
components of $A'$ and $S'$ in $m_t$ units for $m_{\Delta_6}=400$\,GeV and
$m_{\Delta_6}=1$\,TeV.}
\begin{tabular}{|l|c|c|c|}
\hline  & $m_{\Delta_6}=400$\,GeV & $m_{\Delta_6}=1$\,TeV\\
\hline
\hline
$|A'_{12}|/m_t$ & $[2.9 \times 10^{-7}, 8.0 \times 10^{-4}]$ & $[1.1
\times 10^{-6}, 8.2 \times 10^{-4}]$\\
$|A'_{13}|/m_t$ & $[3.7 \times 10^{-2}, 2.5 \times 10^{-1}]$ & $[1.8
\times 10^{-2}, 2.5 \times 10^{-1}]$\\
$|A'_{23}|/m_t$ & $ [1.4 \times 10^{-5}, 5.6 \times 10^{-4}]$ & $ [1.4
\times 10^{-5}, 5.1 \times 10^{-4}]$\\
\hline
\hline
$|S'_{11}|/m_t$ & $[2.2 \times 10^{-9}, 4.0 \times 10^{-6}]$& $[5.3 \times
10^{-9}, 3.9 \times 10^{-6}]$\\
$|S'_{12}|/m_t$ & $[2.6 \times 10^{-7}, 8.1 \times 10^{-4}]$& $[6.5 \times
10^{-7}, 8.2 \times 10^{-4}]$\\
$|S'_{13}|/m_t$ & $[3.7 \times 10^{-2}, 2.5 \times 10^{-1}]$& $[1.8 \times
10^{-2}, 2.5 \times 10^{-1}]$\\
$|S'_{22}|/m_t$ & $[5.9 \times 10^{-5}, 1.7 \times 10^{-3}]$& $[1.0 \times
10^{-4}, 	1.6 \times 10^{-3}]$\\
$|S'_{23}|/m_t$ & $[1.7 \times 10^{-5}, 2.2 \times 10^{-3}]$& $[3.2 \times
10^{-5}, 2.1 \times 10^{-3}]$\\
$|S'_{33}|/m_t$ & $[1.2 \times 10^{-3}, 4.9 \times 10^{-1}]$& $[3.3 \times
10^{-3}, 5.0 \times 10^{-1}]$\\
\hline
\end{tabular}
\end{table}

To summarize, we have shown that the phenomenological constraints on
the form of $A'$ put limitations on allowed form of $S'$, i.e., the
symmetric contribution to the up-quark masses. In fact, $|S'_{12}|$,
$|S'_{13}|$, $|S'_{33}|$ and $|S'_{11}|/|S'_{22}|$ are tied to the
value of one parameter only, i.e, $|U_{31}|$ (or, equivalently,
$v_{45}$). We have shown corresponding ranges of values of $|A'|$ and
$|S'|$ elements in Table~\ref{tab:table1} for all possible
$v_{45}$. Clearly, $v_{45}$ cannot be determined although its upper
bound drops as a function of $m_{\Delta_6}$. This bound will merge
with the lower bound on $v_{45}$ for a rather large value of
$m_{\Delta_6}$, when $\Delta_6$ is out of reach of accelerator
experiments and $|g_6^{13}|$ becomes too large to be trusted. However,
additional constraints on the form of Yukawa couplings in the charged
lepton and down-quark sector would further reduce the available
parameter space for $v_{45}$. If and when $v_{45}$ is better known, we
would be able to bracket $S'$ elements within a narrower range of
values.

Note that $\Delta_6$, being a part of $45$-dimensional
representation of $SU(5)$, is also a part of a $120$-dimensional
representation of
$SO(10)$~\cite{Mohapatra:1979nn,Slansky:1981yr} that is frequently
used to generate antisymmetric Yukawa contributions to charged fermion
masses. There, in the $SO(10)$ framework, the symmetry dictates that
the antisymmetric contributions to charged leptons, down-quarks and
up-quarks are all proportional to the one and the same underlying Yukawa
coupling matrix that, on the other hand, is proportional to $g_6$. In such a
setup one might have additional constraints that could prove sufficient
enough to
pinpoint the exact Yukawa structure. In fact, the lopsided structures
within the $SO(10)$ framework are known to
connect small angles in the quark sector with large angles in the leptonic
sector~\cite{Babu:1995hr} in a natural way. We leave this issue for the
future publication.

\section{Summary}
\label{summary}
We have investigated flavor constraints and predictions in the
presence of the light colored weak singlet scalar, i.e.,
$\Delta_6=(\bar 3, 1, 4/3)$, that couples antisymmetrically, via
Yukawa coupling matrix $g_6$, to the up-quark sector. A tree-level
exchange of $\Delta_6$ contributes to $t\bar t$ production
cross-section in the $u$-channel and, when below $1$\,TeV, can thus
enhance the SM prediction of the forward-backward asymmetry
$A_{FB}^{t\bar t}$ while not altering the production cross-section
$\sigma_{t\bar t}$. This relates the strength of $g_6^{13}$ to the
$\Delta_6$ mass via $ |g_6^{13}| = 0.9 + 2.5 (m_{\Delta_6}/1\e{TeV})$.

Flavor processes that are sensitive to $\Delta_6$ exchange offer a way
to place bounds on the remaining two couplings in $g_6$.  Namely, the
box diagram consisting of $\Delta_6$ and $t$-quark exchanges mediates
\ddbar transitions that allow for bound extraction of $g_6^{23}$. The
bound is phase dependent as summarized in Fig.~\ref{fig:g23} but,
$|g_6^{23}|$ cannot exceed $3.8\E{-3}$ for $m_{\Delta_6}=1$\,TeV. The
CDF search for resonances in the invariant mass spectrum of di-jets as
well as the single top production cross-section measurements at
the Tevatron put bounds on the coupling $g_6^{12}$. We show the
relevant bounds on $|g_6^{12}|$ in Fig.~\ref{fig:bounds}.

Having obtained the upper bounds on $g_6^{12}$ and $g_6^{23}$ we have
also assessed the prospects of looking for radiative top quark decays
at the LHC. For $\Delta_6$ masses below $1\e{TeV}$ scale we have found
the maximal branching fractions of $t \to c \gamma$ and $t \to c G$ of
the order of $10^{-9}$, whereas $t \to u \gamma$ and $t \to u G$
branching fractions are suppressed by an additional 6 orders of
magnitude, owing to strong constraints coming from \ddbar mixing.

These bounds offer an opportunity to constrain the up-quark Yukawa
sector if the weak singlet color scalar has a GUT origin. We have
addressed these constraints within a general $SU(5)$ framework that
employs one $5$- and one $45$-dimensional Higgs representation.  Since
$g_6$ is tied to the antisymmetric contribution of the Higgs doublet
in the $45$-dimensional representation we have deduced the form of the
symmetric Yukawa couplings of the $5$-dimensional Higgs representation
to the matter. We have first described generic features of and correlations
between the symmetric Yukawa couplings that are then confirmed
numerically by using a particular---theoretically well-motivated---GUT
model. We have finally specified the allowed ranges of both symmetric and
antisymmetric Yukawa couplings in units of $m_t$ within that
framework. Our setup is unique since the low-energy phenomenological
constraints allow us to specify Yukawa couplings in the up-quark
sector with great accuracy at the scale of unification.

\begin{acknowledgments}
  This work is supported in part by the European Commission RTN
  network, Contract No. MRTN-CT-2006-035482 (FLAVIAnet) and partially
  by the Slovenian Research Agency. I.D.\ would like to thank Jo\v zef
  Stefan Institute for hospitality where part of this work was
  completed.
\end{acknowledgments}

\bibliography{refs}

\begin{thebibliography}{54}
\expandafter\ifx\csname natexlab\endcsname\relax\def\natexlab#1{#1}\fi
\expandafter\ifx\csname bibnamefont\endcsname\relax
  \def\bibnamefont#1{#1}\fi
\expandafter\ifx\csname bibfnamefont\endcsname\relax
  \def\bibfnamefont#1{#1}\fi
\expandafter\ifx\csname citenamefont\endcsname\relax
  \def\citenamefont#1{#1}\fi
\expandafter\ifx\csname url\endcsname\relax
  \def\url#1{\texttt{#1}}\fi
\expandafter\ifx\csname urlprefix\endcsname\relax\def\urlprefix{URL }\fi
\providecommand{\bibinfo}[2]{#2}
\providecommand{\eprint}[2][]{\url{#2}}

\bibitem[{\citenamefont{Golowich et~al.}(2009)\citenamefont{Golowich, Hewett,
  Pakvasa, and Petrov}}]{Golowich:2009ii}
\bibinfo{author}{\bibfnamefont{E.}~\bibnamefont{Golowich}},
  \bibinfo{author}{\bibfnamefont{J.~A.} \bibnamefont{Hewett}},
  \bibinfo{author}{\bibfnamefont{S.}~\bibnamefont{Pakvasa}}, \bibnamefont{and}
  \bibinfo{author}{\bibfnamefont{A.~A.} \bibnamefont{Petrov}},
  \bibinfo{journal}{Phys. Rev.} \textbf{\bibinfo{volume}{D79}},
  \bibinfo{pages}{114030} (\bibinfo{year}{2009}), \eprint{0903.2830}.

\bibitem[{\citenamefont{Falk et~al.}(2004)\citenamefont{Falk, Grossman, Ligeti,
  Nir, and Petrov}}]{Falk:2004wg}
\bibinfo{author}{\bibfnamefont{A.~F.} \bibnamefont{Falk}},
  \bibinfo{author}{\bibfnamefont{Y.}~\bibnamefont{Grossman}},
  \bibinfo{author}{\bibfnamefont{Z.}~\bibnamefont{Ligeti}},
  \bibinfo{author}{\bibfnamefont{Y.}~\bibnamefont{Nir}}, \bibnamefont{and}
  \bibinfo{author}{\bibfnamefont{A.~A.} \bibnamefont{Petrov}},
  \bibinfo{journal}{Phys. Rev.} \textbf{\bibinfo{volume}{D69}},
  \bibinfo{pages}{114021} (\bibinfo{year}{2004}), \eprint{hep-ph/0402204}.

\bibitem[{\citenamefont{Gedalia et~al.}(2009)\citenamefont{Gedalia, Grossman,
  Nir, and Perez}}]{Gedalia:2009kh}
\bibinfo{author}{\bibfnamefont{O.}~\bibnamefont{Gedalia}},
  \bibinfo{author}{\bibfnamefont{Y.}~\bibnamefont{Grossman}},
  \bibinfo{author}{\bibfnamefont{Y.}~\bibnamefont{Nir}}, \bibnamefont{and}
  \bibinfo{author}{\bibfnamefont{G.}~\bibnamefont{Perez}},
  \bibinfo{journal}{Phys. Rev.} \textbf{\bibinfo{volume}{D80}},
  \bibinfo{pages}{055024} (\bibinfo{year}{2009}), \eprint{0906.1879}.

\bibitem[{\citenamefont{Bigi et~al.}(2009)\citenamefont{Bigi, Blanke, Buras,
  and Recksiegel}}]{Bigi:2009df}
\bibinfo{author}{\bibfnamefont{I.~I.} \bibnamefont{Bigi}},
  \bibinfo{author}{\bibfnamefont{M.}~\bibnamefont{Blanke}},
  \bibinfo{author}{\bibfnamefont{A.~J.} \bibnamefont{Buras}}, \bibnamefont{and}
  \bibinfo{author}{\bibfnamefont{S.}~\bibnamefont{Recksiegel}},
  \bibinfo{journal}{JHEP} \textbf{\bibinfo{volume}{07}}, \bibinfo{pages}{097}
  (\bibinfo{year}{2009}), \eprint{0904.1545}.

\bibitem[{\citenamefont{Grossman et~al.}(2007)\citenamefont{Grossman, Kagan,
  and Nir}}]{Grossman:2006jg}
\bibinfo{author}{\bibfnamefont{Y.}~\bibnamefont{Grossman}},
  \bibinfo{author}{\bibfnamefont{A.~L.} \bibnamefont{Kagan}}, \bibnamefont{and}
  \bibinfo{author}{\bibfnamefont{Y.}~\bibnamefont{Nir}},
  \bibinfo{journal}{Phys. Rev.} \textbf{\bibinfo{volume}{D75}},
  \bibinfo{pages}{036008} (\bibinfo{year}{2007}), \eprint{hep-ph/0609178}.

\bibitem[{\citenamefont{Bigi}(2009)}]{Bigi:2009aw}
\bibinfo{author}{\bibfnamefont{I.~I.} \bibnamefont{Bigi}}
  (\bibinfo{year}{2009}), \eprint{0907.2950}.

\bibitem[{\citenamefont{Petrov}(2010)}]{Petrov:2010gy}
\bibinfo{author}{\bibfnamefont{A.~A.} \bibnamefont{Petrov}}
  (\bibinfo{year}{2010}), \eprint{1003.0906}.

\bibitem[{\citenamefont{Bernreuther}(2008)}]{Bernreuther:2008ju}
\bibinfo{author}{\bibfnamefont{W.}~\bibnamefont{Bernreuther}},
  \bibinfo{journal}{J. Phys.} \textbf{\bibinfo{volume}{G35}},
  \bibinfo{pages}{083001} (\bibinfo{year}{2008}), \eprint{0805.1333}.

\bibitem[{\citenamefont{Aguilar-Saavedra}(2010)}]{AguilarSaavedra:2010rx}
\bibinfo{author}{\bibfnamefont{J.~A.} \bibnamefont{Aguilar-Saavedra}},
  \bibinfo{journal}{Nucl. Phys.} \textbf{\bibinfo{volume}{B837}},
  \bibinfo{pages}{122} (\bibinfo{year}{2010}), \eprint{1003.3173}.

\bibitem[{\citenamefont{Dorsner et~al.}(2010)\citenamefont{Dorsner, Fajfer,
  Kamenik, and Kosnik}}]{Dorsner:2009mq}
\bibinfo{author}{\bibfnamefont{I.}~\bibnamefont{Dorsner}},
  \bibinfo{author}{\bibfnamefont{S.}~\bibnamefont{Fajfer}},
  \bibinfo{author}{\bibfnamefont{J.~F.} \bibnamefont{Kamenik}},
  \bibnamefont{and} \bibinfo{author}{\bibfnamefont{N.}~\bibnamefont{Kosnik}},
  \bibinfo{journal}{Phys. Rev.} \textbf{\bibinfo{volume}{D81}},
  \bibinfo{pages}{055009} (\bibinfo{year}{2010}), \eprint{0912.0972}.

\bibitem[{\citenamefont{Georgi and Glashow}(1974)}]{Georgi:1974sy}
\bibinfo{author}{\bibfnamefont{H.}~\bibnamefont{Georgi}} \bibnamefont{and}
  \bibinfo{author}{\bibfnamefont{S.~L.} \bibnamefont{Glashow}},
  \bibinfo{journal}{Phys. Rev. Lett.} \textbf{\bibinfo{volume}{32}},
  \bibinfo{pages}{438} (\bibinfo{year}{1974}).

\bibitem[{\citenamefont{Fileviez~Perez}(2007)}]{Perez:2007rm}
\bibinfo{author}{\bibfnamefont{P.}~\bibnamefont{Fileviez~Perez}},
  \bibinfo{journal}{Phys. Lett.} \textbf{\bibinfo{volume}{B654}},
  \bibinfo{pages}{189} (\bibinfo{year}{2007}), \eprint{hep-ph/0702287}.

\bibitem[{\citenamefont{Arnold et~al.}(2010)\citenamefont{Arnold, Pospelov,
  Trott, and Wise}}]{Arnold:2009ay}
\bibinfo{author}{\bibfnamefont{J.~M.} \bibnamefont{Arnold}},
  \bibinfo{author}{\bibfnamefont{M.}~\bibnamefont{Pospelov}},
  \bibinfo{author}{\bibfnamefont{M.}~\bibnamefont{Trott}}, \bibnamefont{and}
  \bibinfo{author}{\bibfnamefont{M.~B.} \bibnamefont{Wise}},
  \bibinfo{journal}{JHEP} \textbf{\bibinfo{volume}{01}}, \bibinfo{pages}{073}
  (\bibinfo{year}{2010}), \eprint{0911.2225}.

\bibitem[{\citenamefont{Shu et~al.}(2010)\citenamefont{Shu, Tait, and
  Wang}}]{Shu:2009xf}
\bibinfo{author}{\bibfnamefont{J.}~\bibnamefont{Shu}},
  \bibinfo{author}{\bibfnamefont{T.~M.~P.} \bibnamefont{Tait}},
  \bibnamefont{and} \bibinfo{author}{\bibfnamefont{K.}~\bibnamefont{Wang}},
  \bibinfo{journal}{Phys. Rev.} \textbf{\bibinfo{volume}{D81}},
  \bibinfo{pages}{034012} (\bibinfo{year}{2010}), \eprint{0911.3237}.

\bibitem[{\citenamefont{Georgi and Jarlskog}(1979)}]{Georgi:1979df}
\bibinfo{author}{\bibfnamefont{H.}~\bibnamefont{Georgi}} \bibnamefont{and}
  \bibinfo{author}{\bibfnamefont{C.}~\bibnamefont{Jarlskog}},
  \bibinfo{journal}{Phys. Lett.} \textbf{\bibinfo{volume}{B86}},
  \bibinfo{pages}{297} (\bibinfo{year}{1979}).

\bibitem[{\citenamefont{Dorsner et~al.}(2009)\citenamefont{Dorsner, Fajfer,
  Kamenik, and Kosnik}}]{Dorsner:2009cu}
\bibinfo{author}{\bibfnamefont{I.}~\bibnamefont{Dorsner}},
  \bibinfo{author}{\bibfnamefont{S.}~\bibnamefont{Fajfer}},
  \bibinfo{author}{\bibfnamefont{J.~F.} \bibnamefont{Kamenik}},
  \bibnamefont{and} \bibinfo{author}{\bibfnamefont{N.}~\bibnamefont{Kosnik}},
  \bibinfo{journal}{Phys. Lett.} \textbf{\bibinfo{volume}{B682}},
  \bibinfo{pages}{67} (\bibinfo{year}{2009}), \eprint{0906.5585}.

\bibitem[{\citenamefont{Babu and Ma}(1984)}]{Babu:1984vx}
\bibinfo{author}{\bibfnamefont{K.~S.} \bibnamefont{Babu}} \bibnamefont{and}
  \bibinfo{author}{\bibfnamefont{E.}~\bibnamefont{Ma}}, \bibinfo{journal}{Phys.
  Lett.} \textbf{\bibinfo{volume}{B144}}, \bibinfo{pages}{381}
  (\bibinfo{year}{1984}).

\bibitem[{\citenamefont{Giveon et~al.}(1991)\citenamefont{Giveon, Hall, and
  Sarid}}]{Giveon:1991zm}
\bibinfo{author}{\bibfnamefont{A.}~\bibnamefont{Giveon}},
  \bibinfo{author}{\bibfnamefont{L.~J.} \bibnamefont{Hall}}, \bibnamefont{and}
  \bibinfo{author}{\bibfnamefont{U.}~\bibnamefont{Sarid}},
  \bibinfo{journal}{Phys. Lett.} \textbf{\bibinfo{volume}{B271}},
  \bibinfo{pages}{138} (\bibinfo{year}{1991}).

\bibitem[{\citenamefont{Dorsner and Fileviez~Perez}(2006)}]{Dorsner:2006dj}
\bibinfo{author}{\bibfnamefont{I.}~\bibnamefont{Dorsner}} \bibnamefont{and}
  \bibinfo{author}{\bibfnamefont{P.}~\bibnamefont{Fileviez~Perez}},
  \bibinfo{journal}{Phys. Lett.} \textbf{\bibinfo{volume}{B642}},
  \bibinfo{pages}{248} (\bibinfo{year}{2006}), \eprint{hep-ph/0606062}.

\bibitem[{\citenamefont{Dorsner and Mocioiu}(2008)}]{Dorsner:2007fy}
\bibinfo{author}{\bibfnamefont{I.}~\bibnamefont{Dorsner}} \bibnamefont{and}
  \bibinfo{author}{\bibfnamefont{I.}~\bibnamefont{Mocioiu}},
  \bibinfo{journal}{Nucl. Phys.} \textbf{\bibinfo{volume}{B796}},
  \bibinfo{pages}{123} (\bibinfo{year}{2008}), \eprint{0708.3332}.

\bibitem[{\citenamefont{Ciuchini et~al.}(1998)}]{Ciuchini:1997bw}
\bibinfo{author}{\bibfnamefont{M.}~\bibnamefont{Ciuchini}}
  \bibnamefont{et~al.}, \bibinfo{journal}{Nucl. Phys.}
  \textbf{\bibinfo{volume}{B523}}, \bibinfo{pages}{501} (\bibinfo{year}{1998}),
  \eprint{hep-ph/9711402}.

\bibitem[{\citenamefont{Gupta et~al.}(1997)\citenamefont{Gupta, Bhattacharya,
  and Sharpe}}]{Gupta:1996yt}
\bibinfo{author}{\bibfnamefont{R.}~\bibnamefont{Gupta}},
  \bibinfo{author}{\bibfnamefont{T.}~\bibnamefont{Bhattacharya}},
  \bibnamefont{and} \bibinfo{author}{\bibfnamefont{S.~R.}
  \bibnamefont{Sharpe}}, \bibinfo{journal}{Phys. Rev.}
  \textbf{\bibinfo{volume}{D55}}, \bibinfo{pages}{4036} (\bibinfo{year}{1997}),
  \eprint{hep-lat/9611023}.

\bibitem[{\citenamefont{Eisenstein et~al.}(2008)}]{:2008sq}
\bibinfo{author}{\bibfnamefont{B.~I.} \bibnamefont{Eisenstein}}
  \bibnamefont{et~al.} (\bibinfo{collaboration}{CLEO}), \bibinfo{journal}{Phys.
  Rev.} \textbf{\bibinfo{volume}{D78}}, \bibinfo{pages}{052003}
  (\bibinfo{year}{2008}), \eprint{0806.2112}.

\bibitem[{\citenamefont{Barberio et~al.}(2008)}]{Barberio:2008fa}
\bibinfo{author}{\bibfnamefont{E.}~\bibnamefont{Barberio}} \bibnamefont{et~al.}
  (\bibinfo{collaboration}{Heavy Flavor Averaging Group})
  (\bibinfo{year}{2008}), \eprint{0808.1297}.

\bibitem[{\citenamefont{Grossman et~al.}(2009)\citenamefont{Grossman, Nir, and
  Perez}}]{Grossman:2009mn}
\bibinfo{author}{\bibfnamefont{Y.}~\bibnamefont{Grossman}},
  \bibinfo{author}{\bibfnamefont{Y.}~\bibnamefont{Nir}}, \bibnamefont{and}
  \bibinfo{author}{\bibfnamefont{G.}~\bibnamefont{Perez}},
  \bibinfo{journal}{Phys. Rev. Lett.} \textbf{\bibinfo{volume}{103}},
  \bibinfo{pages}{071602} (\bibinfo{year}{2009}), \eprint{0904.0305}.

\bibitem[{\citenamefont{Fajfer et~al.}(2007)\citenamefont{Fajfer, Kosnik, and
  Prelovsek}}]{Fajfer:2007dy}
\bibinfo{author}{\bibfnamefont{S.}~\bibnamefont{Fajfer}},
  \bibinfo{author}{\bibfnamefont{N.}~\bibnamefont{Kosnik}}, \bibnamefont{and}
  \bibinfo{author}{\bibfnamefont{S.}~\bibnamefont{Prelovsek}},
  \bibinfo{journal}{Phys. Rev.} \textbf{\bibinfo{volume}{D76}},
  \bibinfo{pages}{074010} (\bibinfo{year}{2007}), \eprint{0706.1133}.

\bibitem[{\citenamefont{Burdman et~al.}(1995)\citenamefont{Burdman, Golowich,
  Hewett, and Pakvasa}}]{Burdman:1995te}
\bibinfo{author}{\bibfnamefont{G.}~\bibnamefont{Burdman}},
  \bibinfo{author}{\bibfnamefont{E.}~\bibnamefont{Golowich}},
  \bibinfo{author}{\bibfnamefont{J.}~\bibnamefont{Hewett}}, \bibnamefont{and}
  \bibinfo{author}{\bibfnamefont{S.}~\bibnamefont{Pakvasa}},
  \bibinfo{journal}{Phys. Rev.} \textbf{\bibinfo{volume}{D52}},
  \bibinfo{pages}{6383} (\bibinfo{year}{1995}), \eprint{hep-ph/9502329}.

\bibitem[{\citenamefont{Burdman et~al.}(2002)\citenamefont{Burdman, Golowich,
  Hewett, and Pakvasa}}]{Burdman:2001tf}
\bibinfo{author}{\bibfnamefont{G.}~\bibnamefont{Burdman}},
  \bibinfo{author}{\bibfnamefont{E.}~\bibnamefont{Golowich}},
  \bibinfo{author}{\bibfnamefont{J.}~\bibnamefont{Hewett}}, \bibnamefont{and}
  \bibinfo{author}{\bibfnamefont{S.}~\bibnamefont{Pakvasa}},
  \bibinfo{journal}{Phys. Rev.} \textbf{\bibinfo{volume}{D66}},
  \bibinfo{pages}{014009} (\bibinfo{year}{2002}), \eprint{hep-ph/0112235}.

\bibitem[{\citenamefont{Aaltonen et~al.}(2009)}]{0812.4036}
\bibinfo{author}{\bibfnamefont{T.}~\bibnamefont{Aaltonen}} \bibnamefont{et~al.}
  (\bibinfo{collaboration}{CDF}), \bibinfo{journal}{Phys. Rev.}
  \textbf{\bibinfo{volume}{D79}}, \bibinfo{pages}{112002}
  (\bibinfo{year}{2009}), \eprint{0812.4036}.

\bibitem[{\citenamefont{Group}(2009)}]{0908.2171}
\bibinfo{author}{\bibfnamefont{T.~E.~W.} \bibnamefont{Group}}
  (\bibinfo{collaboration}{CDF}) (\bibinfo{year}{2009}), \eprint{0908.2171}.

\bibitem[{\citenamefont{Lai et~al.}(2000)}]{Lai:1999wy}
\bibinfo{author}{\bibfnamefont{H.~L.} \bibnamefont{Lai}} \bibnamefont{et~al.}
  (\bibinfo{collaboration}{CTEQ}), \bibinfo{journal}{Eur. Phys. J.}
  \textbf{\bibinfo{volume}{C12}}, \bibinfo{pages}{375} (\bibinfo{year}{2000}),
  \eprint{hep-ph/9903282}.

\bibitem[{\citenamefont{Amsler et~al.}(2008)}]{Amsler:2008zzb}
\bibinfo{author}{\bibfnamefont{C.}~\bibnamefont{Amsler}} \bibnamefont{et~al.}
  (\bibinfo{collaboration}{Particle Data Group}), \bibinfo{journal}{Phys.
  Lett.} \textbf{\bibinfo{volume}{B667}}, \bibinfo{pages}{1}
  (\bibinfo{year}{2008}).

\bibitem[{\citenamefont{Kluge et~al.}(2007)\citenamefont{Kluge, Rabbertz, and
  Wobisch}}]{FastNLO}
\bibinfo{author}{\bibfnamefont{T.}~\bibnamefont{Kluge}},
  \bibinfo{author}{\bibfnamefont{K.}~\bibnamefont{Rabbertz}}, \bibnamefont{and}
  \bibinfo{author}{\bibfnamefont{M.}~\bibnamefont{Wobisch}}
  (\bibinfo{year}{2007}), \bibinfo{note}{in Proceedings of the 14th
  International Workshop on Deep Inelastic Scattering (DIS 2006), Tsukuba,
  Japan, 2006}, \eprint{hep-ph/0609285}.

\bibitem[{\citenamefont{Stump et~al.}(2003)}]{CTEQ61}
\bibinfo{author}{\bibfnamefont{D.}~\bibnamefont{Stump}} \bibnamefont{et~al.},
  \bibinfo{journal}{JHEP} \textbf{\bibinfo{volume}{10}}, \bibinfo{pages}{046}
  (\bibinfo{year}{2003}), \eprint{hep-ph/0303013}.

\bibitem[{\citenamefont{Aaltonen et~al.}(2008)}]{FBAexp1}
\bibinfo{author}{\bibfnamefont{T.}~\bibnamefont{Aaltonen}} \bibnamefont{et~al.}
  (\bibinfo{collaboration}{CDF}), \bibinfo{journal}{Phys. Rev. Lett.}
  \textbf{\bibinfo{volume}{101}}, \bibinfo{pages}{202001}
  (\bibinfo{year}{2008}), \eprint{0806.2472}.

\bibitem[{\citenamefont{Abazov et~al.}(2008)}]{FBAexp2}
\bibinfo{author}{\bibfnamefont{V.~M.} \bibnamefont{Abazov}}
  \bibnamefont{et~al.} (\bibinfo{collaboration}{D0}), \bibinfo{journal}{Phys.
  Rev. Lett.} \textbf{\bibinfo{volume}{100}}, \bibinfo{pages}{142002}
  (\bibinfo{year}{2008}), \eprint{0712.0851}.

\bibitem[{\citenamefont{CDF}(2009)}]{FBAexp3}
\bibinfo{author}{\bibnamefont{CDF}}, \bibinfo{journal}{public note 9724}
  (\bibinfo{year}{2009}).

\bibitem[{\citenamefont{Abazov et~al.}(2010)}]{1006.3575}
\bibinfo{author}{\bibfnamefont{V.~M.} \bibnamefont{Abazov}}
  \bibnamefont{et~al.} (\bibinfo{collaboration}{D0}) (\bibinfo{year}{2010}),
  \eprint{1006.3575}.

\bibitem[{\citenamefont{Zhang et~al.}(2009)}]{Zhang:2008yn}
\bibinfo{author}{\bibfnamefont{J.~J.} \bibnamefont{Zhang}}
  \bibnamefont{et~al.}, \bibinfo{journal}{Phys. Rev. Lett.}
  \textbf{\bibinfo{volume}{102}}, \bibinfo{pages}{072001}
  (\bibinfo{year}{2009}), \eprint{0810.3889}.

\bibitem[{\citenamefont{Drobnak et~al.}(2010)\citenamefont{Drobnak, Fajfer, and
  Kamenik}}]{Drobnak:2010wh}
\bibinfo{author}{\bibfnamefont{J.}~\bibnamefont{Drobnak}},
  \bibinfo{author}{\bibfnamefont{S.}~\bibnamefont{Fajfer}}, \bibnamefont{and}
  \bibinfo{author}{\bibfnamefont{J.~F.} \bibnamefont{Kamenik}},
  \bibinfo{journal}{Phys. Rev. Lett.} \textbf{\bibinfo{volume}{104}},
  \bibinfo{pages}{252001} (\bibinfo{year}{2010}), \eprint{1004.0620}.

\bibitem[{\citenamefont{Zhang et~al.}(2010)}]{Zhang:2010bm}
\bibinfo{author}{\bibfnamefont{J.~J.} \bibnamefont{Zhang}} \bibnamefont{et~al.}
  (\bibinfo{year}{2010}), \eprint{1004.0898}.

\bibitem[{\citenamefont{Minkowski}(1977)}]{Minkowski:1977sc}
\bibinfo{author}{\bibfnamefont{P.}~\bibnamefont{Minkowski}},
  \bibinfo{journal}{Phys. Lett.} \textbf{\bibinfo{volume}{B67}},
  \bibinfo{pages}{421} (\bibinfo{year}{1977}).

\bibitem[{\citenamefont{Yanagida}(1979)}]{Yanagida:1979as}
\bibinfo{author}{\bibfnamefont{T.}~\bibnamefont{Yanagida}}
  (\bibinfo{year}{1979}), \bibinfo{note}{in Proceedings of the Workshop on the
  Baryon Number of the Universe and Unified Theories, Tsukuba, Japan, 13-14 Feb
  1979}.

\bibitem[{\citenamefont{Gell-Mann et~al.}(1980)\citenamefont{Gell-Mann, Ramond,
  and Slansky}}]{GellMann:1980vs}
\bibinfo{author}{\bibfnamefont{M.}~\bibnamefont{Gell-Mann}},
  \bibinfo{author}{\bibfnamefont{P.}~\bibnamefont{Ramond}}, \bibnamefont{and}
  \bibinfo{author}{\bibfnamefont{R.}~\bibnamefont{Slansky}}
  (\bibinfo{year}{1980}), \bibinfo{note}{print-80-0576 (CERN)}.

\bibitem[{\citenamefont{Glashow}(1980)}]{Glashow:1979nm}
\bibinfo{author}{\bibfnamefont{S.~L.} \bibnamefont{Glashow}},
  \bibinfo{journal}{NATO Adv. Study Inst. Ser. B Phys.}
  \textbf{\bibinfo{volume}{59}}, \bibinfo{pages}{687} (\bibinfo{year}{1980}).

\bibitem[{\citenamefont{Mohapatra and Senjanovic}(1980)}]{Mohapatra:1979ia}
\bibinfo{author}{\bibfnamefont{R.~N.} \bibnamefont{Mohapatra}}
  \bibnamefont{and}
  \bibinfo{author}{\bibfnamefont{G.}~\bibnamefont{Senjanovic}},
  \bibinfo{journal}{Phys. Rev. Lett.} \textbf{\bibinfo{volume}{44}},
  \bibinfo{pages}{912} (\bibinfo{year}{1980}).

\bibitem[{\citenamefont{Foot et~al.}(1989)\citenamefont{Foot, Lew, He, and
  Joshi}}]{Foot:1988aq}
\bibinfo{author}{\bibfnamefont{R.}~\bibnamefont{Foot}},
  \bibinfo{author}{\bibfnamefont{H.}~\bibnamefont{Lew}},
  \bibinfo{author}{\bibfnamefont{X.~G.} \bibnamefont{He}}, \bibnamefont{and}
  \bibinfo{author}{\bibfnamefont{G.~C.} \bibnamefont{Joshi}},
  \bibinfo{journal}{Z. Phys.} \textbf{\bibinfo{volume}{C44}},
  \bibinfo{pages}{441} (\bibinfo{year}{1989}).

\bibitem[{\citenamefont{Ma}(1998)}]{Ma:1998dn}
\bibinfo{author}{\bibfnamefont{E.}~\bibnamefont{Ma}}, \bibinfo{journal}{Phys.
  Rev. Lett.} \textbf{\bibinfo{volume}{81}}, \bibinfo{pages}{1171}
  (\bibinfo{year}{1998}), \eprint{hep-ph/9805219}.

\bibitem[{\citenamefont{Bajc and Senjanovic}(2007)}]{Bajc:2006ia}
\bibinfo{author}{\bibfnamefont{B.}~\bibnamefont{Bajc}} \bibnamefont{and}
  \bibinfo{author}{\bibfnamefont{G.}~\bibnamefont{Senjanovic}},
  \bibinfo{journal}{JHEP} \textbf{\bibinfo{volume}{08}}, \bibinfo{pages}{014}
  (\bibinfo{year}{2007}), \eprint{hep-ph/0612029}.

\bibitem[{\citenamefont{Dorsner and Fileviez~Perez}(2007)}]{Dorsner:2006fx}
\bibinfo{author}{\bibfnamefont{I.}~\bibnamefont{Dorsner}} \bibnamefont{and}
  \bibinfo{author}{\bibfnamefont{P.}~\bibnamefont{Fileviez~Perez}},
  \bibinfo{journal}{JHEP} \textbf{\bibinfo{volume}{06}}, \bibinfo{pages}{029}
  (\bibinfo{year}{2007}), \eprint{hep-ph/0612216}.

\bibitem[{\citenamefont{Bajc et~al.}(2007)\citenamefont{Bajc, Nemevsek, and
  Senjanovic}}]{Bajc:2007zf}
\bibinfo{author}{\bibfnamefont{B.}~\bibnamefont{Bajc}},
  \bibinfo{author}{\bibfnamefont{M.}~\bibnamefont{Nemevsek}}, \bibnamefont{and}
  \bibinfo{author}{\bibfnamefont{G.}~\bibnamefont{Senjanovic}},
  \bibinfo{journal}{Phys. Rev.} \textbf{\bibinfo{volume}{D76}},
  \bibinfo{pages}{055011} (\bibinfo{year}{2007}), \eprint{hep-ph/0703080}.

\bibitem[{\citenamefont{Mohapatra and Sakita}(1980)}]{Mohapatra:1979nn}
\bibinfo{author}{\bibfnamefont{R.~N.} \bibnamefont{Mohapatra}}
  \bibnamefont{and} \bibinfo{author}{\bibfnamefont{B.}~\bibnamefont{Sakita}},
  \bibinfo{journal}{Phys. Rev.} \textbf{\bibinfo{volume}{D21}},
  \bibinfo{pages}{1062} (\bibinfo{year}{1980}).

\bibitem[{\citenamefont{Slansky}(1981)}]{Slansky:1981yr}
\bibinfo{author}{\bibfnamefont{R.}~\bibnamefont{Slansky}},
  \bibinfo{journal}{Phys. Rept.} \textbf{\bibinfo{volume}{79}},
  \bibinfo{pages}{1} (\bibinfo{year}{1981}).

\bibitem[{\citenamefont{Babu and Barr}(1996)}]{Babu:1995hr}
\bibinfo{author}{\bibfnamefont{K.~S.} \bibnamefont{Babu}} \bibnamefont{and}
  \bibinfo{author}{\bibfnamefont{S.~M.} \bibnamefont{Barr}},
  \bibinfo{journal}{Phys. Lett.} \textbf{\bibinfo{volume}{B381}},
  \bibinfo{pages}{202} (\bibinfo{year}{1996}), \eprint{hep-ph/9511446}.

\end{thebibliography}

\end{document}